%% file: main.tex
\documentclass[conference]{IEEEtran}
\IEEEoverridecommandlockouts
\usepackage{cite}
\usepackage{amsmath,amssymb,amsfonts}
\usepackage{algorithmic}
\usepackage{graphicx}
\usepackage{textcomp}
\usepackage{xcolor}

\usepackage[ruled, linesnumbered]{algorithm2e}
\SetKwInOut{Input}{Input}
\SetKwInOut{Output}{Output}
\SetKwComment{Comment}{/* }{ */}
\usepackage{amsthm}         
\usepackage{caption}        
\usepackage{hyperref}       
\usepackage{xspace}         

\def\BibTeX{{\rm B\kern-.05em{\sc i\kern-.025em b}\kern-.08em
    T\kern-.1667em\lower.7ex\hbox{E}\kern-.125emX}}

\newcommand{\sysname}{\mbox{Pylon}\xspace}
\newcommand{\plainsysname}{Pylon\xspace}
\newcommand{\fasttext}{\textit{fastText}\xspace}
\newcommand{\wte}{\textit{WTE}\xspace}
\newcommand{\bert}{\textit{BERT}\xspace}
\newcommand{\dddl}{$D^{3}L$\xspace}
\newcommand{\ourfasttext}{\mbox{\textit{\sysname}-\fasttext}\xspace}
\newcommand{\ourwte}{\mbox{\textit{\sysname}-\wte}\xspace}
\newcommand{\ourlm}{\mbox{\textit{\sysname-LM}}\xspace}

\theoremstyle{definition}
\newtheorem{definition}{Definition}

\begin{document}
\title{\plainsysname: Semantic Table Union Search in Data Lakes 
}

\author{\IEEEauthorblockN{Tianji Cong}
\IEEEauthorblockA{\textit{University of Michigan}\\
Ann Arbor, USA \\
congtj@umich.edu}
\and
\IEEEauthorblockN{Fatemeh Nargesian}
\IEEEauthorblockA{\textit{University of Rochester}\\
Rochester, USA \\
fnargesian@rochester.edu}
\and
\IEEEauthorblockN{H. V. Jagadish}
\IEEEauthorblockA{\textit{University of Michigan}\\
Ann Arbor, USA \\
jag@umich.edu}
}

\maketitle

\input{sections/00_abstract}


\input{sections/01_introduction}
\input{sections/02_background}
\input{sections/03_methodology}
\input{sections/04_experiments}
\input{sections/05_relatedWork}
\input{sections/06_conclusion}


\bibliographystyle{IEEEtran}
\bibliography{sections/07_reference}

\end{document}

%% file: sections/00_abstract.tex
\begin{abstract}
  The large size and fast growth of data repositories, such as data lakes, has spurred the need for data discovery to help analysts find related data. The problem has become challenging as (i) a user typically does not know what datasets exist in an enormous data repository; and (ii) there is usually a lack of a unified data model to capture the interrelationships between heterogeneous datasets from disparate sources. In this work, we address one important class of discovery needs: finding union-able tables.
  
  The task is to find tables in a data lake that can be unioned with a given query table. The challenge is to recognize union-able columns even if they are represented differently. In this paper, we propose a data-driven learning approach: specifically, an unsupervised representation learning and embedding retrieval task. Our key idea is to exploit self-supervised contrastive learning to learn an embedding model that takes into account the indexing/search data structure and produces embeddings close by for columns with semantically similar values while pushing apart columns with semantically dissimilar values. We then find union-able tables based on similarities between their constituent columns in embedding space. On a real-world data lake, we demonstrate that our best-performing model achieves significant improvements in precision ($16\% \uparrow$), recall ($17\% \uparrow $), and query response time (7x faster) compared to the state-of-the-art.
\end{abstract}

%% file: sections/01_introduction.tex
\section{Introduction}
  Recent years have witnessed a vast growth in the amount of data available to the public, particularly from data markets, open data portals, and data communities (e.g., Wikidata and Kaggle)~\cite{chapman2020dataset}. To benefit from the many new opportunities for data analytics and data science, the user first usually has to find related datasets in a large repository (e.g., data lake). Common practice in production is to provide a keyword search interface over the metadata of datasets~\cite{brickley2019google} but users often have discovery needs that cannot be precisely expressed by keywords. The challenge for a system is to support users with varying discovery needs, without the help of a unified data model capturing the interrelationships between datasets.
  
  In response to the challenge, there are many ongoing efforts under the umbrella of data discovery. One task of interest in data discovery is to find union-able tables~\cite{Cafarella08, Sarma12, Nargesian18, bogatu2020dataset} with the aim of adding additional relevant rows to a user-provided table. Figure~\ref{fig:unionable_table_example} shows an example of two tables union-able over four pairs of attributes. In general, the literature considers two tables union-able if they share attributes from the same domain and assumes the union-ability of two attributes can be implied by some notion of similarity. We refer to the problem of finding union-able tables as table union search (as termed in ~\cite{Nargesian18}) in the rest of the paper.

\begin{table*}[htbp!]
\centering
\resizebox{\textwidth}{!}{%
\begin{tabular}{rrrrr}
\hline
\multicolumn{1}{c}{id} &
  \multicolumn{1}{c}{title} &
  \multicolumn{1}{c}{authors} &
  \multicolumn{1}{c}{venue} &
  \multicolumn{1}{c}{year} \\ \hline
671167 &
  A Database System for Real-Time Event Aggregat... &
  Jerry Baulier, Stephen Blott, Henry F. Korth, ... &
  Very Large Data Bases &
  1998 \\
672964 &
  Integrating a Structured-Text Retrieval System... &
  Tak W. Yan, Jurgen Annevelink &
  Very Large Data Bases &
  1994 \\
872823 & Evaluating probabilistic queries over imprecis... & Reynold Cheng, Dmitri V. Kalashnikov, Sunil Pr... & International Conference on Management of Data & 2003 \\
\multicolumn{1}{c}{...} &
  \multicolumn{1}{c}{...} &
  \multicolumn{1}{c}{...} &
  \multicolumn{1}{c}{...} &
  \multicolumn{1}{c}{...} \\ \hline
\end{tabular}%
}
\end{table*}

\begin{table*}[ht!]
\centering
\resizebox{\textwidth}{!}{%
\begin{tabular}{cccccc}
\hline
Title &
  Authors &
  Platform &
  Cited\_url &
  Cited\_count &
  Year \\ \hline
Fg-index: towards verification-free query proc... &
  J Cheng, Y Ke, W Ng, A Lu &
  Proceedings of the 2007 ACM SIGMOD internation... &
  https://scholar.google.com/scholar?oi=bibs\&hl=... &
  286 &
  2007 \\
Efficient query processing on graph databases &
  J Cheng, Y Ke, W Ng &
  ACM Transactions on Database Systems (TODS) 34... &
  https://scholar.google.com/scholar?oi=bibs\&hl=... &
  83 &
  2009 \\
Context-aware object connection discovery in l... &
  J Cheng, Y Ke, W Ng, JX Yu &
  2009 IEEE 25th International Conference on Dat... &
  https://scholar.google.com/scholar?oi=bibs\&hl=... &
  66 &
  2009 \\
... &
  ... &
  ... &
  ... &
  ... &
  ... \\ \hline
\end{tabular}%
}
\captionof{figure}{An example of two tables union-able over four pairs of attributes: title - Title, authors - Authors, venue - Platform, and year - Year.}
\label{fig:unionable_table_example}
\end{table*}
  
  The typical solution path is to first identify union-able attributes (or columns in the tables) with the aid of an indexing/search data structure (e.g., locality sensitive hashing) and then aggregate column-level results to obtain candidate union-able tables. To uncover the union-ability of attributes, both syntactic and semantic methods have been exploited. Syntactic methods are the easiest, and have been used the longest. While they are robust at catching small changes, such as capitalization or the use of a hyphen, they are unable to address even the use of common synonyms. Semantic methods offer the possibility of finding union-able columns of semantically similar values despite their syntactic dissimilarity (e.g., ``venue" column and ``platform" column in figure~\ref{fig:unionable_table_example}). ~\cite{Sarma12, Nargesian18} link cell values to entity classes in an external ontology and compare similarity of entity sets. ~\cite{Nargesian18, bogatu2020dataset} use off-the-shelf word embeddings to measure semantics. Both methods have notable limitations. \cite{Nargesian18} observed that only $13\%$ of attribute values of their collected Open Data tables can be mapped to entities in YAGO~\cite{suchanek2007yago}, one of the largest and publicly available ontologies. Although word embeddings can provide more semantic coverage of attributes, they are subject to the training text corpus and may not generalize well to textual data in tables~\cite{Gunther21, koutras2021valentine}. Recent advance in tabular embeddings~\cite{deng2020turl, tang2021rpt} may mitigate the issue, however, we argue that these models in training do not take into account the indexing/search data structure, which is a core component of any efficient solution to table union search. In particular, ~\cite{Nargesian18, bogatu2020dataset} rely on the correlation between column union-ability and cosine similarity of their embeddings. Nevertheless, the embeddings they use and recent more advanced tabular models are not optimized in training for approximate cosine similarity search so the performance of indexing/search data structure is suboptimal.

  Instead of relying on low-coverage ontologies or pre-trained embeddings that are unaware of required data structure in table union search, we propose a data-driven learning approach to capture data semantics and model characteristics of the essential indexing/search data structure. We also argue that the popular classification formulation in the literature~\cite{deng2020turl, li2020deep, li2021deep, tang2021rpt} is less feasible for table union search. On one hand, there is no large-scale labeled dataset for table union search. The only publicly available benchmark~\cite{Nargesian18} with table- and column-level ground truth contains a limited number of tables synthesized from only 32 base tables, which is far from being enough for training purposes. On the other hand, even if the training data problem were resolved, we would only be able to determine column matches pairwise. Unlike tasks (e.g., semantic column type annotation and entity matching) that can be formulated as a classification problem, finding union-able tables is a search problem. It would be very inefficient to exhaustively consider every query column and every column in the data lake pairwise to predict union-ability. In short, the inherent search nature of the problem makes the supervised classification formulation infeasible and it is important to jointly consider representation learning and the indexing/search data structure.


In this work, we overcome the aforementioned difficulties by casting table union search as an unsupervised representation learning and embedding retrieval task. Our goal is to learn column-level embeddings into a high-dimensional feature space that also models characteristics of the indexing/search data structure. Locality search in this feature space can then directly be used for union-able table search. To achieve this goal, our key idea is to exploit self-supervised contrastive learning to learn an embedding model that produces embeddings with high similarity measure (which is used in indexing/search data structure) for columns with semantically similar values and pushes away columns with semantically dissimilar values. We propose \sysname, a novel self-supervised contrastive learning framework that learns column representations and serves table union search without relying on labeled data.
  
  There are two main challenges in the development of \sysname: 

\begin{enumerate}
    
\item How to learn embeddings that capture both data semantics and model characteristics of indexing/search data structure? Existing embedding models may capture semantics from a large training corpus but they are ignorant of additional data structure necessary in table union search. In other words, the potential of locality search is not fully realized with existing embeddings.
    
\item How to create training data without human labeling? The self-supervised contrastive learning technique requires constructing positive and negative examples from data themselves. In the field of computer vision where contrastive learning first took off, ~\cite{chen2020simple} applies a series of random data augmentation of crop, flip, color jitter, and grayscale to generate stochastic views of an image. These views preserve the semantic class label of the image and so make positive examples for training. They further consider any two views not from the same image as a negative example. However, the tabular data modality is dramatically different from images and it remains unclear how to create different views of tables while keeping the semantics.
  \end{enumerate}
  
  In summary, we make the following contributions:
  \begin{itemize}
    \item We formulate semantic table union search as an unsupervised representation learning and embedding retrieval problem, and propose to use self-supervised contrastive learning to model both data semantics and indexing/search data structure.
    \item We present \sysname, a contrastive learning framework for learning semantic column representations from large collections of tables without using labels. In particular, we propose two simple yet effective strategies to construct training data in an unsupervised manner. 
    \item We empirically show that our approach is both more effective and efficient than existing embedding methods on a real-world data lake of GitHub tables and a synthetic public benchmark of open data tables. On the GitHub data lake, two of our model variants outperform their corresponding baseline version by $14\%$ and $6\%$ respectively on both precision and recall. We also observe that they speed up the query response time by 2.7x and 9x respectively. We (plan to) open-source our new benchmark of GitHub tables for future research study.
    \item We demonstrate that our embedding approach can be further augmented by syntactic measures and that our best ensemble model has significant advantages over the state-of-the-art (namely, $D^{3}L$~\cite{bogatu2020dataset}), more than $15\%$ improvement in precision and recall, and 7x faster in query response time.
  \end{itemize}

  We give a formal problem setup and background about embedding models in Section 2. We describe our framework \sysname including embedding training and search in Section 3. Section 4 reports experiments that validate our approach. We discuss related work in Section 5 and conclude in Section 6.

%% file: sections/02_background.tex
\section{Problem Definition \& Background}
   In this section, we start by describing the formal problem setup in~\ref{subsec:table_union_search}, and then provide an overview of how table union search is different from a well-established data management problem, namely schema matching, and a related problem of join discovery in~\ref{subsec:pos_wrt_sm_and_jd}. We finally elaborate on the challenges of applying representation learning for table union search in~\ref{subsec:rl_challenges_in_tus}.
  
\subsection{Table Union Search}\label{subsec:table_union_search}
  The table union search problem~\cite{Nargesian18} is motivated by the need to augment a (target) table at hand with additional data from other tables containing similar information. For example, starting with a table about traffic accidents in one state for a particular year, an analyst may wish to find similar traffic accident data for other states and years. Ideally, these tables would have the same schema (e.g. data from the same state agency for two different years) so that we could simply union the row-sets. However, this is typically not the case for data recorded independently (e.g. data from different states). We consider two tables union-able if they share attributes from the same domain. Also, as in prior work on this topic, we assume the union-ability of attributes can be quantified by some notion of similarity.
    
  \begin{definition}[Attribute Union-ability]\label{def:attribute_unionability}
    Given two attributes $A$ and $B$, the attribute union-ability $\mathcal{U}_{attr}(A, B)$ is defined as $$\mathcal{U}_{attr}(A, B) = \mathcal{M}(\mathcal{T}(A), \mathcal{T}(B))$$ where $\mathcal{T}(\cdot)$ is a feature extraction technique that transforms raw columns (attribute names, attribute values, or both) to a feature space and $\mathcal{M}(\cdot, \cdot)$ is a similarity measure between two instances in the feature space.
  \end{definition}
    
  With the definition of attribute union-ability, we can define table uniona-bility as a bipartite graph matching problem where the disjoint sets of vertices are attributes of the target table and the source table respectively, and edges can be defined by attribute union-ability. In this paper, we restrict ourselves to the class of greedy solutions. Therefore, we formalize the definition  table union-ability as a greedy matching problem as follows:
    
  \begin{definition}[Union-able Tables]\label{def:union-able_tables}
    A source table $S$ with attributes $\mathcal{B}=\{B_{j}\}_{j=1}^{n}$ is union-able to a target table $T$ with attributes $\mathcal{A}=\{A_{i}\}_{i=1}^{m}$ if there exists a one-to-one mapping $g: \mathcal{A'} (\neq \emptyset) \subseteq \mathcal{A} \rightarrow \mathcal{B'} \subseteq \mathcal{B}$ such that
    \begin{enumerate}
      \item $|\mathcal{A'}| = |\mathcal{B'}|$;
      \item $\forall A_{i} \in \mathcal{A'}$, $\mathcal{U}_{attr}(A_{i}, g(A_{i})) \geq \tau$ where $$g(A_{i}) = \displaystyle \arg \max_{B_{j}} \{\mathcal{U}_{attr}(A_{i}, B_{j}): 1 \leq j \leq n\}$$ and $\tau$ is a pre-defined similarity threshold.
    \end{enumerate}
  \end{definition}
    
  \begin{definition}[Table Union-ability]\label{def:table_unionability}
    Following notations in Definition~\ref{def:union-able_tables}, the table union-ability $\mathcal{U}(S, T)$ is defined as $$\mathcal{U}(S, T) = \frac{\sum_{i=1}^{l} w_{i} \cdot \mathcal{U}_{attr}(A_{i}, g(A_{i}))}{\sum_{i=1}^{l} w_{i}}$$ where $l$ is the number of union-able attribute pairs between the target table $T$ and a source table $S$, and $w_{i}$ weights the contribution of the attribute pair $(A_{i}, g(A_{i}))$ to the table union-ability.
  \end{definition}
  
  Considering the scale of the dataset repository, we also follow the common practice\cite{Sarma12, Nargesian18, bogatu2020dataset} of performing top-$k$ search. The table union search problem is formally defined as below.
    
  \begin{definition}[Top-$k$ Table Union Search]
    Given a table corpus $\mathcal{S}$, a target table $T$, and a constant $k$, find up to $k$ candidate tables $S_{1}, S_{2}, ..., S_{k} \in \mathcal{S}$ in descending order of table union-ability with respect to the query table $T$ such that $S_{1}, S_{2}, ..., S_{k}$ are most likely to be union-able with $T$.
  \end{definition}

\subsection{How Table Union Search Differs from Schema Matching and Join Discovery?}\label{subsec:pos_wrt_sm_and_jd}
  Despite of overlapping elements, we emphasize the complexity of table union search over related problems of schema matching and join discovery. As a long-standing and well-studied problem in data integration, schema matching takes a pair of schemas as input and returns a mapping between columns of two schemas that semantically correspond to each other~\cite{DBLP:journals/vldb/RahmB01}. Conceptually, table union search can be viewed as an extreme extension of schema matching. Instead of having two schemas as inputs, table union search only has one while having to search another (partially) matching schema in a large corpus (e.g., data lakes) in the first place. Essentially, aside from the matching component, table union search needs to address the additional problem of identifying matching candidates among many non-matching ones, which is a significantly more challenging setup. Similarly, fuzzy join~\cite{DBLP:journals/pvldb/ChenWNC19, DBLP:journals/pvldb/SuriIRR21} assumes a restrictive setup with a pair of input datasets. In their experiments, the second dataset in the pair is usually a syntactically perturbed variant of the first dataset and thus cannot mimic the complexity of data lakes consisting of heterogeneous datasets across domains.

  A problem more related to table union search is join discovery~\cite{fernandez2018aurum, zhu2019josie, bogatu2020dataset}, which targets at finding joinable tables in data lakes. Both are search problems, nevertheless, table union search needs to ideally identify a matching between blocks of columns in two tables as opposed to identifying a pair of join keys between two tables in join discovery. This difference poses higher demand for embedding quality in table union search because moderately high similarity between every pair of columns makes it harder to match union-able pairs.

\subsection{General Challenges}\label{subsec:rl_challenges_in_tus}
  Representation learning for tables has achieved excellent results for many table-centric tasks. We hypothesize that the table union search problem can also benefit from advances in table modeling. However, several challenges remain to be addressed.
  \begin{enumerate}
    \item To the best of our knowledge, no prior work has taken the learning approach for table union search. We argue that this is mainly because neither the supervised learning setting nor the popular pre-training and fine-tuning paradigm is directly applicable for the problem. It is inefficient to formulate the underlying search of union-able columns as a classification problem. In a supervised learning setting, one can attempt to train a classifier to predict whether two columns are union-able, but it will quickly become computationally prohibitive in the search phase to classify every pair of target column in a query table with every column in a large corpus.
    \item The scarcity of table union search datasets is another severe bottleneck of applying a learning approach and studying the problem in general. The only publicly available benchmark~\cite{Nargesian18} with table- and column-level ground truth is synthesized from only 32 base tables, which is barely enough for evaluation. It is also very laborious and time consuming to label such datasets, as curators need to examine every pair of columns for every pair of tables in a collection.
    \item Efficient solutions to table union search involve two stages: profiling (e.g., embed columns into a feature space) and index-based search. Taking off-the-shelf embedding models or training a new model without considering the indexing/search data structure indispensable in table union search is suboptimal. We argue that aligning representation learning with indexing/search data structure can further improve effectiveness and efficiency of a solution to table union search.
  \end{enumerate}
    
  In the next section, we present our design that contributes a representation learning approach to table union search while effectively addressing the challenges we point out here.

%% file: sections/03_methodology.tex
\begin{figure*}[ht!]
  \centering
  \includegraphics[width=0.95\textwidth]{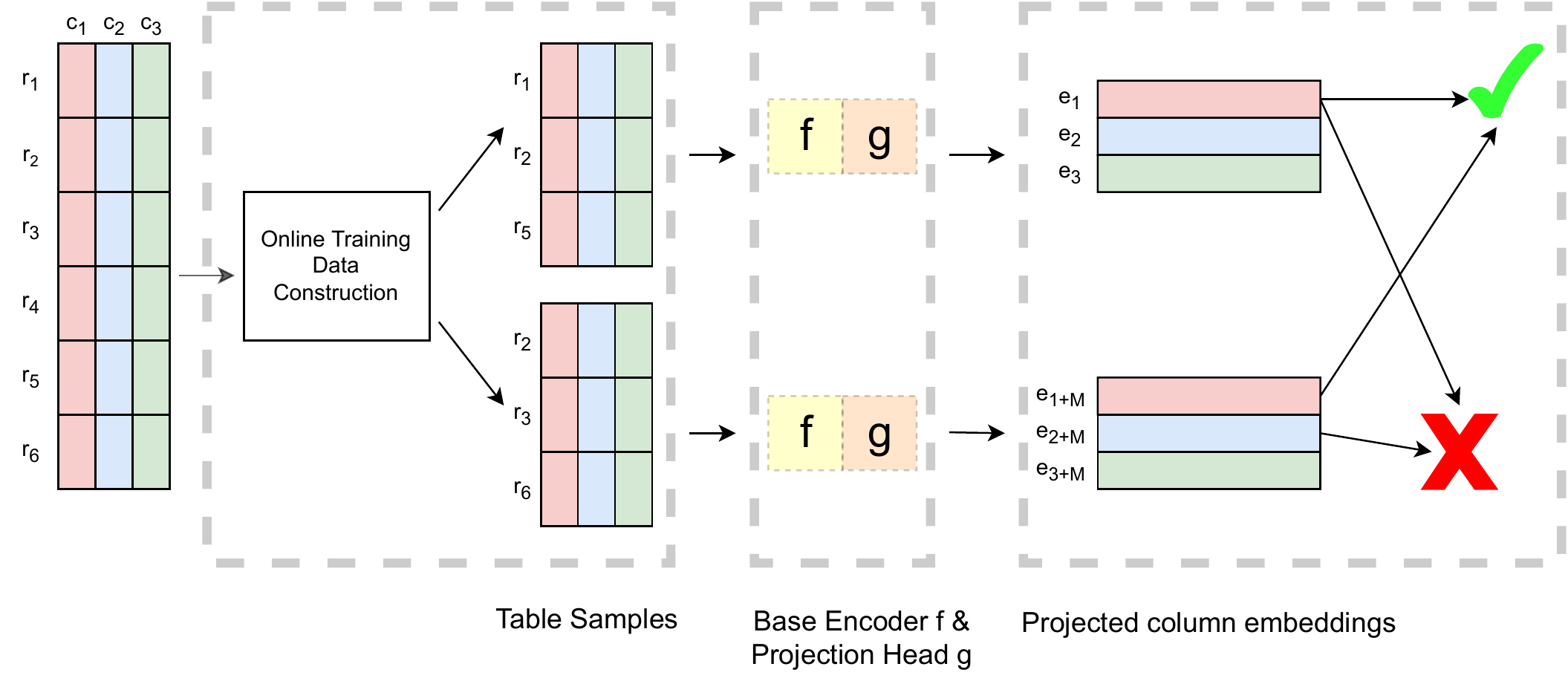}
  \caption{Training workflow of \sysname (with online training data construction).}
  \label{fig:cl_workflow}
\end{figure*}

\section{Pylon: A Self-Supervised Contrastive Learning Framework for Tabular Data}
  Our key idea is to exploit self-supervised contrastive learning that can provide a feasible training objective for learning effective column representations for the table union search problem while not requiring any labeled data and taking into account the indexing/search data structure (corresp. to challenge 1 and 3). Within the framework of contrastive learning, we propose two strategies that arithmetically construct training data from unlabeled data to tackle challenge 2. 
  
\subsection{Contrastive Learning}
  The high-level goal of contrastive learning is to learn to distinguish (so called "contrast") between pairs of similar and dissimilar instances. Ideally, in the learned representation space, similar instances stay close to each other whereas dissimilar ones are pushed far away. A pair of instances is considered similar and labeled a positive example in training if it comprises different views of the same object; otherwise, they are considered dissimilar and make a negative example. Contrastive learning has been used extensively in computer vision~\cite{chen2020simple}, where a positive example consists of a pair of augmented images transformed from the same image (e.g., by applying cropping or color distortion).
  
  We introduce, \sysname, our self-supervised contrastive learning framework for table union search. As table union search begins by finding union-able columns, \sysname is designed to generate a vector representation for each column of input tables where columns containing semantically similar values have embeddings closer to one another.
  
\subsection{\sysname Workflow}
  Figure~\ref{fig:cl_workflow} shows the training workflow of the framework that consists of the following major components.
    
  \textbf{Training Data Construction.} Without labeled data, the success of contrastive learning hinges on the construction of positive and negative examples from data themselves. To make positive examples, it requires an operation to transform a data instance in a way that introduces variations while preserving the semantics. As table union search builds on union-able column search, we propose two strategies to construct positive and negative examples at the column level.

  \begin{enumerate}
    \item \textit{Online sampling strategy.} Consider a training batch of $N$ tables $\{T_{i}\}_{i=1}^{N}$ where each table $T_{i}$ has $m_{i}$ columns $\{C^{i}_{j}\}_{j=1}^{m_{i}}$, giving $M = \sum_{i=1}^{N} m_{i}$ columns in total. We obtain a positive example of column pairs $(x_{k}, x_{k+M})$ ($1 \leq k \leq M$) by randomly sampling values from each column $C^{i}_{j}$ of each table $T_{i}$. Since both $x_{k}$ and $x_{k+M}$ are samples from the same column, we consider they share semantics and make a positive example. The sampling process yields $2M$ column instances, and we treat the other $2(M-1)$ samples as negatives with respect to $x_{k}$. In other words, considering $(x_{k}, x_{k+M})$ and $(x_{k+M}, x_{k})$ as distinct positive examples, we construct $2M$ positive examples and $2M(M-1)$ negative examples from each training batch.
    \smallskip
    \item \textit{Offline approximate matching strategy.} An alternative is to construct positive examples ahead of training. Instead of relying on ad-hoc sampling, we can leverage existing approaches to find a union-able candidate for each column, which in turn makes positive examples in training. Based on the observation that top-$k$ union-able column search of existing techniques is reasonably precise when $k=1$, we are able to use this approximate matching without human involvement. We find it produces valid results and models do not deteriorate due to potential false positives in training data.
  \end{enumerate}
    
  \textbf{Base Encoder \& Projection Head.} We pass column instances $\{x_{k}\}_{k=1}^{2M}$ through a base encoder $f(\cdot)$ to get initial column embeddings $\{e_{k}\}_{k=1}^{2M}$. Note that our contrastive learning framework is flexible about the choice of the base encoder. The encoder can give embeddings at token/cell/column level, and if necessary, we can apply aggregation (e.g, average) to obtain column-level embeddings. Our framework has the flexibility to benefit from the advance of modeling techniques over time without being tied to a specific model. We describe the choices of $f(\cdot)$ we experiment with in subsection \ref{subsec:base_encoder_choices}.
    
  Following the encoder, a small multi-layer neural network $g(\cdot)$, called projection head, maps the representations from the encoder to a latent space through linear transformations and non-linear activation in between. Note that unlike the practice in CV which discards projection head in inference and uses encoder outputs for downstream tasks, we preserve projection head and use projected embeddings for table union search. This is because we found projected embeddings yield better performance in initial experiments, and for encoders like word embedding models, only projection head is trainable and has to be preserved for inference. For simplicity, we keep using the notations $\{e_{k}\}_{k=1}^{2M}$ for projection outputs.

  \textbf{Contrastive Loss.} The training objective is a core component in the framework, which drives the learned representations towards the characteristics of the indexing/search data structure in table union search. For example, considering an indexing/search data structure that approximates cosine similarity of vectors, the model should ideally learn to produce embeddings with high cosine similarity for union-able columns in our case.
  
  One common setting of contrastive learning defines a prediction task of identifying positive examples from the training batch. Given embedded columns $\{e_{k}\}_{k=1}^{2M}$, the model learns to predict $e_{k+M}$ as the most similar one to $e_{k}$ and vice versa for each $e_{k} (1\leq k \leq M)$. Here, one can choose a similarity measure that aligns with locality search. In other words, depending on what similarity measure for which the indexing/search data structure is designed, we can push that similarity measure into model learning. As an illustration, the similarity between any two instances $e_{i}$ and $e_{j}$ can be measured by their cosine similarity as $$sim(i, j) = \frac{e_{i}^{T} \; e_{j}}{\lVert e_{i} \rVert \lVert e_{j} \rVert}$$ and the loss is calculated as $$l(k, k+M) = -\log \frac{\exp{(sim(k, k+M)} \;/\; \tau)}{\sum_{l=1, l \neq k}^{2M} \exp{(sim(k, l) \;/\; \tau )}}$$ where $\tau > 0$ is a scaling hyper-parameter called temperature. Minimizing $l(k, k+M)$ is equivalent to maximizing the probability of $e_{k+M}$ being the most similar to $e_{k}$ in terms of cosine similarity among all the embedded columns except $e_{k}$ itself. In this way, the learned column representations are more feasible to be used with the indexing/search data structure that approximates cosine similarity compared to embeddings given by any other models that are not trained to optimize for this purpose (e.g., tabular language models that are trained with the objective to recover masked tokens or entities~\cite{deng2020turl}).

  Finally, the loss over all the $2M$ positive column pairs in a training batch is computed as $$L = \frac{1}{2M} \sum_{k=1}^{M} [l(k, k+M), l(k+M, k)].$$ Algorithm~\ref{alg:pylon_training} summarizes the training process of contrastive learning in \sysname. 

\subsection{Choices of the Base Encoder}\label{subsec:base_encoder_choices}
  Although we expect the input to the contrastive loss function to be column embeddings, the base encoder does not necessarily need to give column embeddings directly. It is possible for the encoder model to generate embeddings at different granularity (i.e., token/cell/column) because we can apply aggregation if necessary. We describe the basic encoding process of embedding models we experimented with in section~\ref{experiments}.
    
  \textbf{Word Embedding Models (WEM).} As a WEM assigns a fix representation to a token, WEM-based encoders treat each column independently as a document where a standard text parser tokenizes data values in a column. With a fastText embedding model, we first get cell embeddings by averaging token embeddings in each cell and then aggregate cell embeddings to get a column embedding. More interestingly, web table embedding models~\cite{Gunther21} consider each cell as a single token (they concatenate tokens in a cell with underscores) and output embeddings at cell level. Nevertheless, we aggregate cell embeddings to derive the column embedding.
    
  \textbf{Language Models (LM).} Since a table is a cohesive structure for storing data, considering values in neighboring columns could integrate context into the embeddings and help mitigate ambiguity in unionable column search. For example, encoding column "year" in figure~\ref{fig:unionable_table_example} individually loses the context that this column refers to the publication year of research papers. In this case, the embeddings of "Year" columns in the corpus are less distinguishable (in terms of cosine similarity) even though they may refer to different concepts of year such as the birth year of people or the release year of movies. With context provided by other columns like "Title" and "Venue", it is more likely that "Year" columns appearing in tables about papers are more close to each other than "Year" columns in tables about other topics, which helps find more related tables.
    
  We leverage LMs to derive contextual column embeddings. We first serialize each row in $T_{i}$ as a sequence by concatenating tokenized cell values. For example, the first row of the table at the top in Figure~\ref{fig:unionable_table_example} will be linearized as follows $$\mbox{\footnotesize [CLS] title $\vert$ A Database $\dots$ [SEP] authors $\vert$ Jerry $\dots$ [SEP] $\dots$ [END]}$$ The sequence is annotated with special tokens in the LM where [CLS] token indicates the beginning of the sequence, [END] token indicates the end, and [SEP] tokens separate cell values in different columns. Then the LM takes in each sequence and generates a contextual representation for each token in the sequence (essentially taking into account the relation between values in the same row). We apply mean pooling to tokens in the same cell and get cell embeddings. To consider the relation of values in the same column, we adopt the vertical attention mechanism in ~\cite{yin2020tabert} to have weighted column embeddings by attending to all of the sampled cells in the same column.

  Word embedding models have previously been used to find union-able tables. Two state-of-the-art choices are \fasttext and \wte (web table embeddings~\cite{Gunther21}). Language models have not thus far been used for the union-ability problem. \bert\cite{devlin2019bert} is a leading language model used for many purposes today. We develop three versions of \sysname, one for each of these three encoder choices: \fasttext, \wte, and a \bert-based language model, and refer to the derived models as \ourfasttext, \ourwte, \ourlm respectively. We evaluate the effect of encoder choices in subsection~\ref{exp:results}.

\subsection{Embedding Indexing and Search}
  \begin{algorithm}[t!]
    \caption{\sysname Contrastive Learning}\label{alg:pylon_training}
    \Input{\quad $\mathcal{S}$, a corpus of tables; \\
           \quad $N$, batch size; \\
           \quad $p$, training hyper-parameters.}
    \Output{\quad $g \circ f$, a \sysname model.}
    \medskip
    $g \circ f \gets \text{initialize\_model()}$\;
    \For{mini-batch of $N$ tables $\{C_{j}^{i}\}_{i=1}^{N}$ from $\mathcal{S}$}{
      $\{x_{k}, x_{k+M}\}_{k=1}^{M} \gets \text{construct\_training\_data}(\{C_{j}^{i}\}_{i=1}^{N})$ \;
      \smallskip
      $\{e_{k}\}_{k=1}^{2M} \gets g \circ f.\text{encode\_and\_embed}(\{x_{k}, x_{k+M}\}_{k=1}^{M})$ \;
      \smallskip
      /* Define $l(k, k+M) = -\log \frac{\exp{(sim(k, k+M)} \;/\; \tau)}{\sum_{l=1, l \neq k}^{2M} \exp{(sim(k, l) \;/\; \tau )}}$ and $sim(i, j) = \frac{e_{i}^{T} \; e_{j}}{\lVert e_{i} \rVert \lVert e_{j} \rVert}$ */ \\
      $\mathcal{L} \gets \frac{1}{2M} \displaystyle\sum_{k=1}^{M} [l(k, k+M), l(k+M, k)]$ \;
      $g \circ f \gets \mathcal{L}.\text{backpropagate}(p)$ \;
    }
    \Return $g \circ f$\;
  \end{algorithm}
    
  To avoid exhaustive comparisons of column embeddings over a large corpus at query time, we use locality-sensitive hashing (LSH)~\cite{indyk1998approximate} for approximate nearest neighbor search and treat union-able column search as an LSH-index lookup task~\cite{Nargesian18, bogatu2020dataset}. Note that the specific choice of indexing/search data structure is flexible (one can use a more advanced technique like Hierarchical Navigable Small World~\cite{malkov2018efficient}). The key is that the similarity measure approximated by the indexing/search data structure should align with the similarity measure employed in the training objective of contrastive learning so that the learned embeddings are optimized for the indexing/search data structure.
  
  LSH utilizes a family of hash functions that maximize collisions for similar inputs. The result of LSH indexing is that similar inputs produce the same hash value and are bucketed together whereas dissimilar inputs are ideally placed in different buckets. For approximate search relative to the cosine similarity, we index all column embeddings in a random projection LSH index~\cite{charikar2002similarity}. The idea of random projection is to separate data points in a high-dimensional vector space by inserting hyper-planes. Embeddings with high cosine similarity tend to lie on the same side of many hyper-planes. 
    
  \begin{algorithm}[t!]
    \caption{Top-$k$ Table Union Search}\label{alg:topk_tus}
    \Input{\quad $\mathcal{I}$, a populated LSH index; \\ $\quad Q$, a query table; \\ $\quad k$, a constant.}
    \Output{\quad top-$k$ union-able tables.}
    \medskip
    $column\_candidates \gets \{\}$\;
    $column\_scores \gets \{\}$\;
    \For{$c \in Q.columns$}{
      $c\_candidates, c\_scores \gets \mathcal{I}.\text{lookup}(c)$\;
      $column\_candidates[c].\text{add}(c\_candidates)$\;
      $column\_scores[c].\text{add}(c\_scores)$\;
    }
      
    $table\_candidates \gets \text{group\_by}(column\_candidates)$\;
    $ranked\_table\_candidates \gets \text{\small cmpt\_table\_unionability}(\textit{\small table\_candidates, column\_scores})$\;
    \Return $ranked\_table\_candidates[:k]$\;
  \end{algorithm}
    
  Algorithm~\ref{alg:topk_tus} summarizes the top-$k$ table union search. Following Definition~\ref{def:attribute_unionability}, we instantiate the union-ability of two attributes as the cosine similarity of their embeddings ($c\_scores$ in line 4 of Algorithm~\ref{alg:topk_tus}). Line 8 groups retrieved column candidates across query columns by their table sources. To decide on the table union-ability from Definition~\ref{def:table_unionability} ($cmpt\_table\_unionability$ in line 9), we use the same weighting strategy as ~\cite{bogatu2020dataset} over query attributes and corresponding matching attributes in candidate tables. For a target attribute $A$, let $R_{A}$ denote the distribution of all similarity (union-ability) scores between $A$ and any attribute $B$ returned by the LSH index. The weight $w$ of a similarity score $\mathcal{U}_{attr}(A, B)$ is given by the cumulative distribution function of $R_{A}$ evaluated at $\mathcal{U}_{attr}(A, B)$: $$w = \Pr(\mathcal{U}_{attr}(A, B') \leq \mathcal{U}_{attr}(A, B)), \forall \; \mathcal{U}_{attr}(A, B') \in R_{A}$$ In other words, a similarity score is weighted by its percentile in the distribution. This weighting scheme helps balance between a candidate table with a few union-able attributes of high similarity scores and another candidate table with more union-able attributes but of lower similarity scores.

  Using the same index and search structure as previous works makes it transparent to compare our embedding approach with theirs in effectiveness and efficiency.

\subsection{Integrating Syntactic Methods}
  Thus far, we have focused purely on semantic methods to unify similar attributes. It makes sense to prefer semantic methods to syntactic ones because of their potential robustness to many different types of variation. However, we note that syntactic methods are based on measures of similarity very different from semantic methods. Intuitively, one should expect to be able to do better if we can integrate the two.

  Indeed, some previous work \cite{Nargesian18, bogatu2020dataset} has made this observation as well, and shown that an ensemble of semantic and syntactic methods can do better than either on its own. The \textit{\sysname} semantic method permits the use of an additional complementary syntactic method. As in \cite{bogatu2020dataset}, we independently obtain scores from the two methods and then use the average of the two as our final score.

%% file: sections/04_experiments.tex
\section{Experiments}\label{experiments}
  We first evaluate the effectiveness and efficiency of three model variants from our contrastive learning framework and compare them with their corresponding base encoders. We then demonstrate that our embedding approach is orthogonal to existing syntactic measures, which can further improve the results. We finally compare our best model with the state-of-the-art \dddl~\cite{bogatu2020dataset}.

\subsection{Datasets and Metrics}
  \textbf{TUS Benchmark.} ~\cite{Nargesian18} compiles the first benchmark with ground truth out of Canadian and UK open data lakes. They synthesize around $5,000$ tables from $32$ base tables by performing random projection and selection. They also generate a smaller benchmark consisting of around $1,500$ tables from $10$ base tables in the same manner. We refer to them as TUS-Large and TUS-Small respectively. 
  
  \textbf{\sysname Benchmark. } We create a new dataset from GitTables~\cite{GitTables}, a data lake of $1.7M$ tables extracted from CSV files on GitHub. The benchmark comprises 1,746 tables including union-able table subsets under topics selected from Schema.org~\cite{guha2016schema}: scholarly article, job posting, and music playlist. We end up with these three topics since we can find a fair number of union-able tables of them from diverse sources in the corpus (we can easily find union-able tables from a single source but they are less interesting for table union search as simple syntactic methods can identify all of them because of the same schema and consistent value representations). 
    
  \textbf{Cleaning and Construction.} We download three largest subsets of GitTables ("object", "thing", and "whole") and preprocess them by removing HTML files, tables without headers, rows with foreign languages, and finally small tables with fewer than four rows or four columns. We cluster the resulting tables by their schema and perform a keyword search over schema with keywords related to three topics. We manually select 35 union-able tables of topic scholarly article, 41 tables of topic job posting, and 48 tables of topic music playlist. We then randomly sample 100,000 tables for training, 5,000 tables for validation, and put the rest of tables as negatives\footnote{we filtered these tables using their schema to reduce the chance of them being union-able to selected tables in the union-able subsets.} in a pool with selected union-able table subsets for the search evaluation. 
    
  Table~\ref{tab:bm_stats} provides an overview of basic table statistics of each evaluation benchmark.

\begin{table}[ht!]
\caption{Basic statistics of evaluation datasets.}
\resizebox{0.95\columnwidth}{!}{%
\begin{tabular}{lccc}
\hline
                         & \textbf{Pylon} & \textbf{TUS-Small} & \textbf{TUS-Large} \\ \hline
\textbf{\# Tables}       & 1,746          & 1,530              & 5,043              \\
\textbf{\# Base Tables}  & 1,746          & 10                 & 32                 \\
\textbf{Avg. \# Rows}    & 115            & 4,466              & 1,915              \\
\textbf{Avg. \# Columns} & 10             & 10                 & 11                 \\
\textbf{\# Queries}      & 124            & 1,327              & 4,296              \\
\textbf{Avg. \# Answers} & 42             & 174                & 280                \\ \hline
\end{tabular}%
}
\label{tab:bm_stats}
\end{table}

  \textbf{Metrics.} For effectiveness, we report both precision and recall of top-$k$ search with varying $k$. At each value of $k$, we average the precision and recall numbers over all the queries. We consider a table candidate a true positive with respect to the target table if it is in labeled ground truth. We do not require perfect attribute pair matching as it is a more challenging setting and requires laborious column-level labeling.
    
  As to efficiency, we report indexing time (i.e., total amount of time in minutes to index all columns in a dataset) and query response time (i.e., average amount of time in seconds for the LSH index to return results over all queries in a dataset\footnote{Current implementations include data loading and logging time as well.}). 
    
  In evaluation, we randomly sample 1000 queries from TUS-Large for efficient experiment purposes. The query subset has an average answer size of 277, which is very close to that of the full query set (i.e., 280). We use all the queries in \sysname and TUS-Small datasets.

\subsection{Baselines}
  We consider two embedding methods and one full approach as baselines for comparison.
    
  \fasttext. Many data management tasks not limited to table union search~\cite{fernandez2018seeping, Nargesian18, cappuzzo2020creating} have adopted \fasttext in their approach, which is a popular word embedding model trained on Wikipedia documents.
    
  \wte.~\cite{Gunther21} devises a word embedding-based technique to represent text values in Web tables. They generate text sequences from tables for training by serializing tables in two different ways that capture row-wise relations and relations between schema and data values respectively. It is reported that the model using both serialization obtained the best performance in a task of ranking union-able columns. 
    
  \dddl.~\cite{bogatu2020dataset} proposed a distance-based framework \dddl that uses five types of evidence to decide on column union-ability: (i) attribute name similarity; (ii) attribute extent overlap; (iii) word-embedding similarity; (iv) format representation similarity; (v) domain distribution similarity for numerical attributes. Their aggregated approach is shown to be more effective and efficient than previous work~\cite{Nargesian18, fernandez2018aurum} on the TUS benchmark and another self-curated dataset of open data tables. 

  \begin{figure*}[!ht]
    \centering 
    \minipage{0.45\textwidth}
      \includegraphics[width=\linewidth]{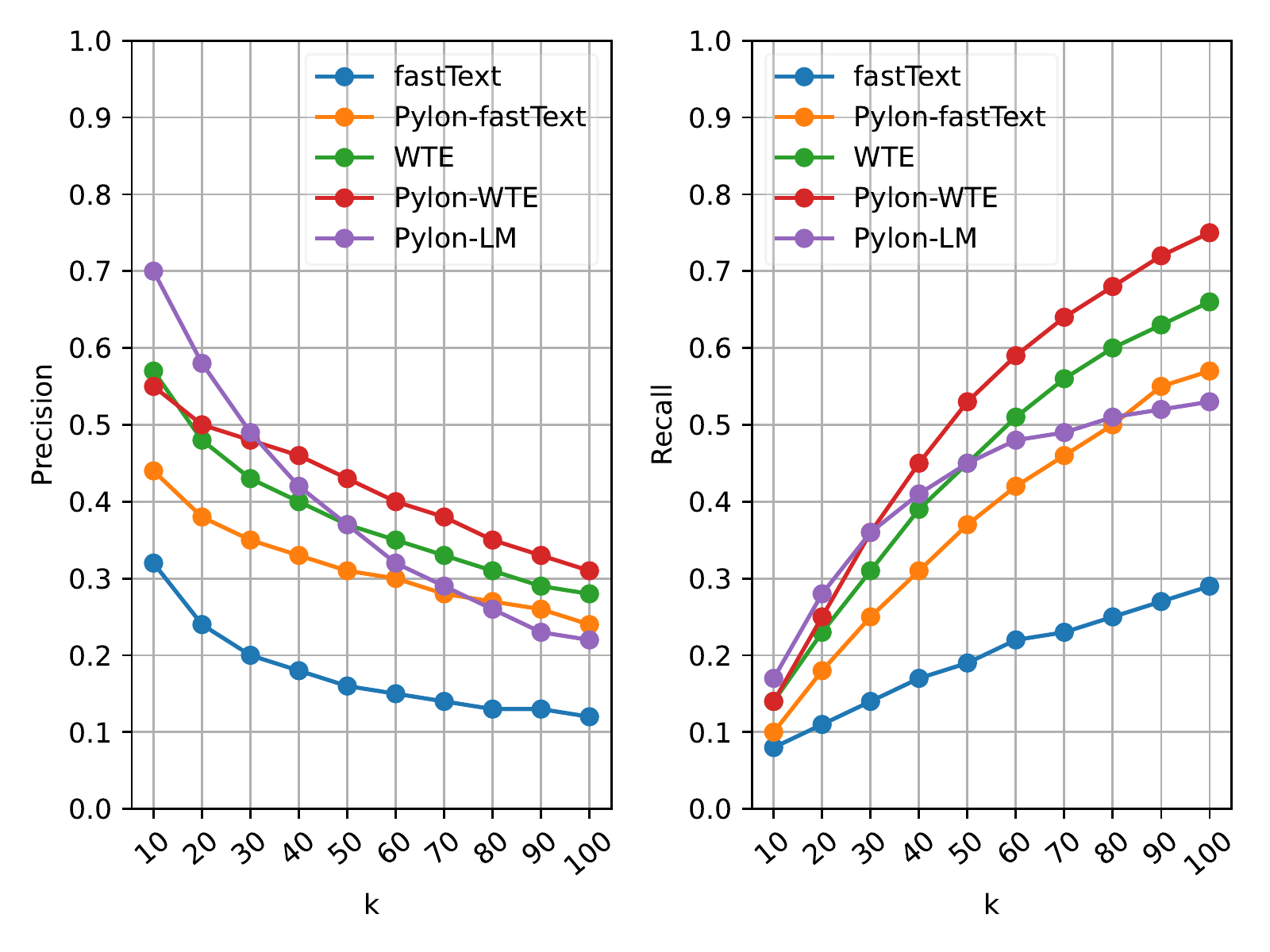}
      \caption{Top-k precision and recall of embedding measures on the \sysname dataset.}
      \label{fig:effectiveness_pylon}
    \endminipage\hfill
    \minipage{0.45\textwidth}
      \includegraphics[width=\linewidth]{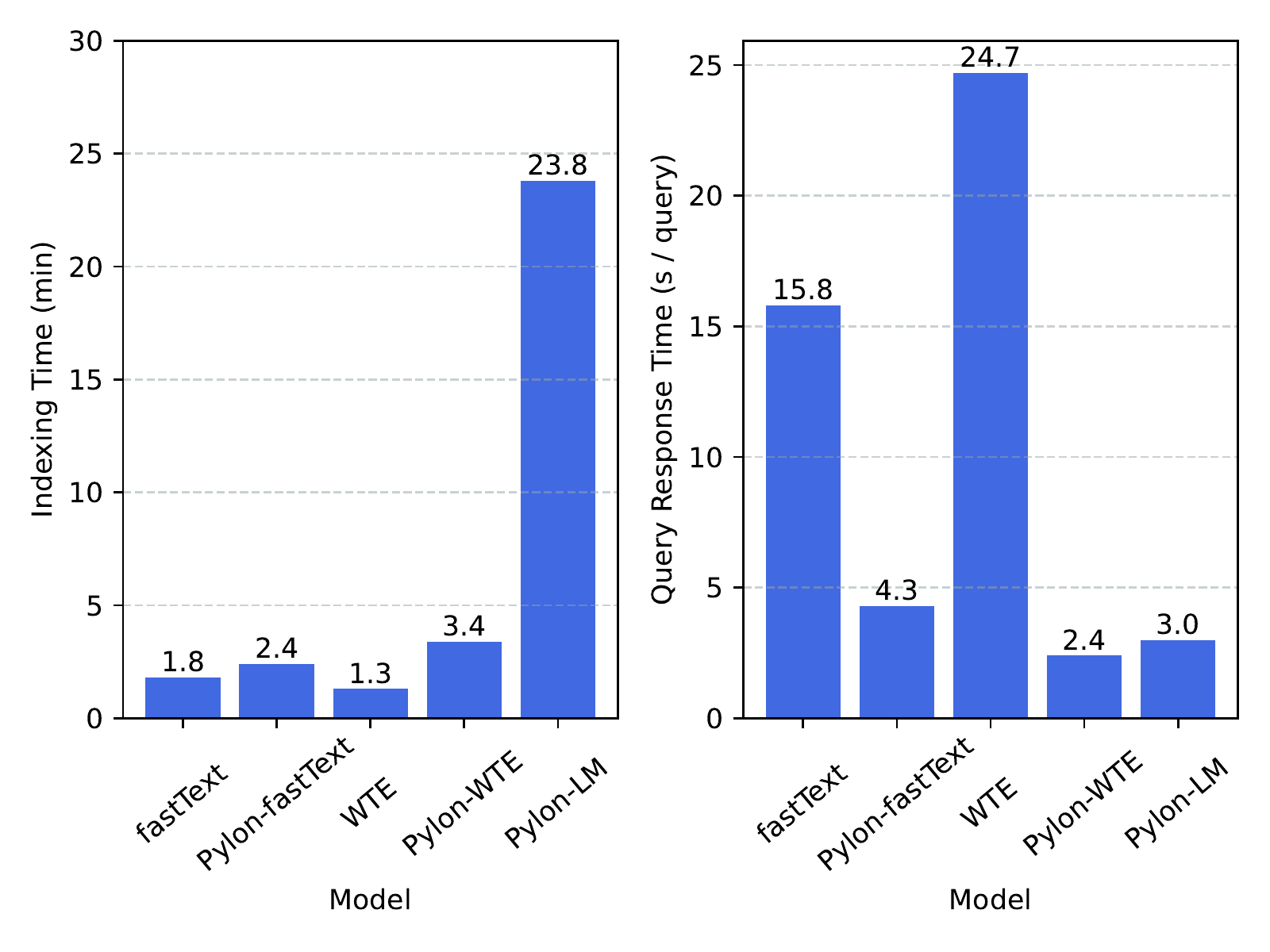}
      \caption{Indexing time and query response time on the \sysname dataset.}
      \label{fig:exp1_pylon_efficiency}
    \endminipage
  \end{figure*}

\subsection{Comparisons of Interest}
  We have 5 variants of \sysname to compare against baseline systems for both effectiveness and efficiency in identifying union-able tables using semantic similarity methods: 3 variants from the online training data construction strategy and 2 variants from the offline data construction strategy. In addition, we have 3 syntactic similarity measures that could be used to augment each of these 5 variants. Finally, we have 3 baselines, two of which are semantic word embedding based, and hence could also be augmented with the syntactic similarity measures. The third baseline ($\mathit{D^{3}L}$) already integrates both syntactic and semantic similarity, and hence does not benefit from additional augmentation with syntactic techniques.

  Since there are a very large number of alternatives to compare, we break up the comparisons into four sets, as follows, and present the results for each set separately. For the first three sets, we restrict ourselves to the online training data construction strategy for \sysname. We refer to the derived models as \ourfasttext, \ourwte, \ourlm respectively based on the corresponding encoder choice. Results for the offline data construction strategy show generally similar trends, and the most interesting are shown in the fourth set.

  The first set of comparisons look purely at semantic methods, considering the 3 variants of \sysname and comparing them to the first two baselines. We leave out \dddl because it already incorporates syntactic methods as well. The second set of comparisons look purely at the benefit obtained when semantic methods are enhanced with syntactic measures. We do so for all methods evaluated in the first set. Finally, we bring everything together by comparing the best methods of the second set with the best integrated baseline, \dddl. This is the final top line "take away" from the experiments, eliding details from the first two sets of comparisons.

  \begin{figure*}[ht!]
    \centering
    \minipage{0.48\linewidth}
      \includegraphics[width=\textwidth]{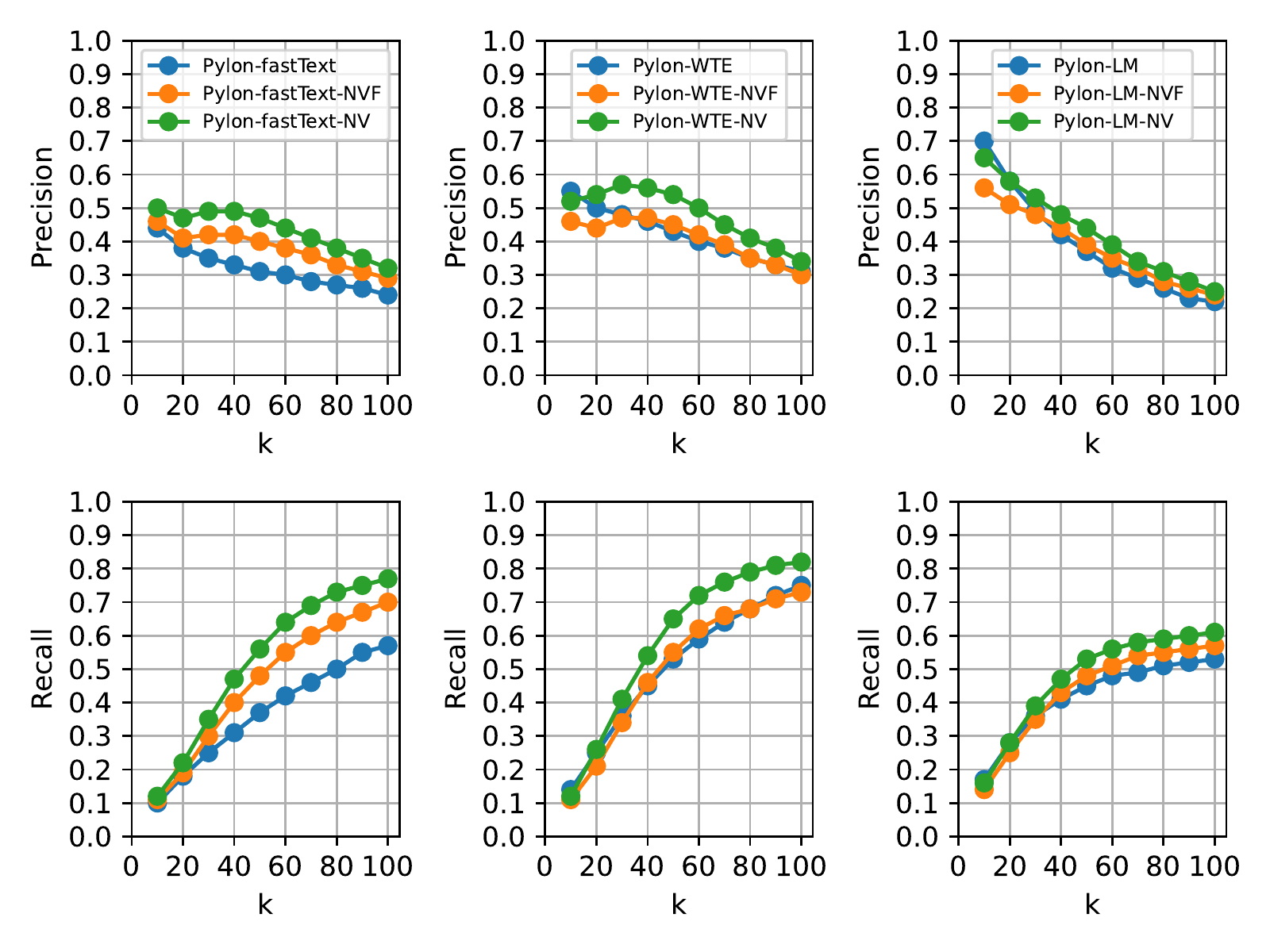}
      \caption*{(a) \sysname dataset}
      \label{fig:exp2_pylon_performance}
      \endminipage\hfill
    \minipage{0.48\linewidth}
      \includegraphics[width=\textwidth]{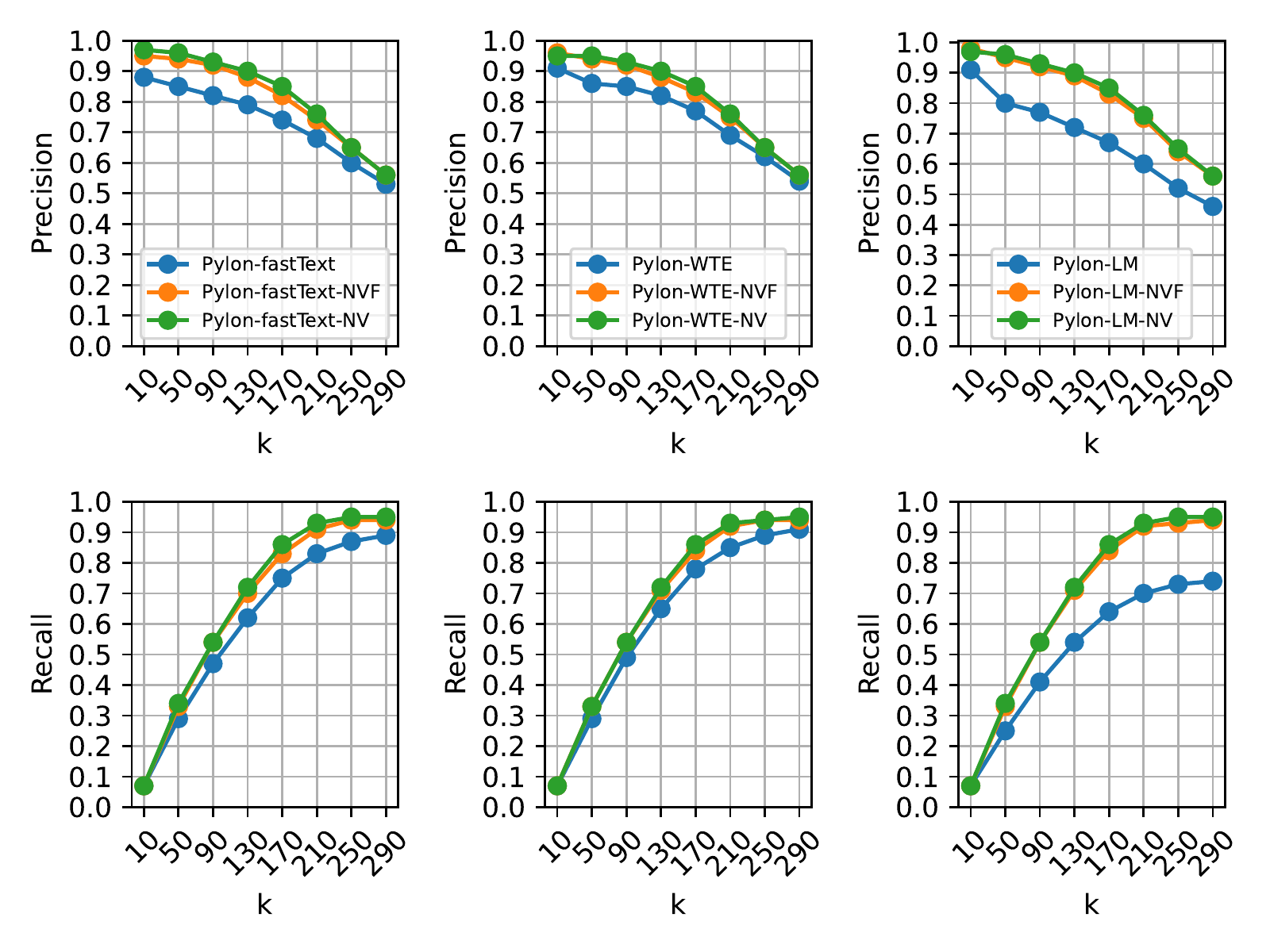}
      \caption*{(b) TUS-Small dataset}
      \label{fig:exp2_tus_small_performance}
    \endminipage
    \caption{Precision and recall (w.r.t. varying $k$) of the ensemble of \sysname embedding models and syntactic measures.}
    \label{fig:exp2_effectiveness}
\end{figure*}

\subsection{Experiment Details}
  As to model training, we train \ourfasttext for 50 epochs with a batch size of 16 on 2 NVIDIA GeForce RTX 2080 Ti GPUs; \ourwte for 20 epochs with a batch size of 32 on a single NVIDIA Tesla P100 GPU; \ourlm for 20 epochs with a batch size of 8 on 4 NVIDIA Tesla P100 GPUs from Google Cloud Platform. As seen in table~\ref{tab:model_training_time}, the training is especially efficient for simple word embedding encoders (as only parameters in projection head are updated) and the offline data construction strategy (as embeddings are pre-computed before training). We save the models with the smallest validation loss. The model training is implemented in PyTorch~\cite{NEURIPS2019_9015} and PyTorch Lightning. 

\begin{table}[h!]
  \centering
  \caption{Model training time (min / epoch) where each model is defined by the encoder choice and the training data construction strategy.}
  \label{tab:model_training_time}
  \resizebox{0.95\columnwidth}{!}{%
  \begin{tabular}{ccc}
  \hline
                          & Online Sampling & Offline Approximate Matching \\ \hline
  \textit{Pylon-fastText} & 6.5             & 0.42                         \\
  \textit{Pylon-WTE}      & 0.99            & 0.13                         \\
  \textit{Pylon-LM}       & 33              & -                            \\
  \hline
  \end{tabular}%
  }
\end{table}
    
  For evaluation of table union search, we set the similarity threshold of LSH index to $0.7$ in all experiments and use the same hash size as \dddl. We run all evaluation on a Ubuntu 20.04.4 LTS machine with 128 GiB RAM and Intel(R) Xeon(R) Bronze 3106 CPU @ 1.70GHz. 

\subsection{Results}\label{exp:results}
  As \sysname is an embedding-based approach, we first evaluate \sysname model variants against embedding baselines \fasttext and \wte, and inspect the effects of contrastive learning on them.
  
  \textbf{Experiment 1(a): Comparison of effectiveness between \sysname model variants and their corresponding base encoders.} Figure~\ref{fig:effectiveness_pylon} shows the precision and recall of each embedding measure on the \sysname benchmark. Both \ourwte and \ourfasttext outperform their corresponding base models with a notable margin. When $k=40$, around the average answer size, \ourwte is $6\%$ better than \wte on both metrics, and \ourfasttext performs better than \fasttext by $15\%$ on precision and $14\%$ on recall. 
    
  Overall, our \ourwte model consistently achieves the highest precision and recall as $k$ increases. We also note that \ourlm has strong performance up until $k=30$ but degrades after that. This is because \ourlm only samples 10 rows from each table to construct embeddings (for indexing and query efficiency) while other word-embedding methods can afford to encode the entire table at low indexing time, which we demonstrate in experiment 1(b).
  
  \textbf{Experiment 1(b): Comparison of efficiency between \sysname model variants and their corresponding base encoders.} In figure~\ref{fig:exp1_pylon_efficiency}, we see both embedding baselines are very efficient in index construction and it takes less than 2 minutes to index the entire \sysname dataset. Unlike fixed embeddings, our models need to infer embeddings at runtime. For \ourfasttext and \ourwte, since the encoder is fixed, the inference cost is exclusively from projection head. It takes both less than 3.5 minutes to build the index. In contrast, the runtime inference cost of \ourlm is more expensive as the language model has much more complex architecture and has 130M parameters versus 35.8K parameters in projection head. We also acknowledge the less efficient implementation of embedding inference at this point (e.g., run inference for each column without batch predictions). Nevertheless, indexing time, as a one-time overhead, can be amortized among queries.
    
  On the other hand, all of our models are considerably more efficient in query response time. \ourfasttext is 2.7x faster than \fasttext and \ourwte is 9x faster than \wte. The significant speedup of query response time is attributed to contrastive learning where embeddings of attribute values occurring in the same context are pushed close to each other and to have high cosine similarity whereas embeddings of two random columns are pushed apart and to have low cosine similarity. As the embedding similarity between two random columns is suppressed, this dramatically reduces the chance of two random columns sharing many LSH buckets. In other words, our column embeddings are trained in a specific way that aligns with the similarity measure LSH index approximates and LSH index can process much fewer candidates at the configured similarity threshold.
    
\begin{figure}[ht!]
    \centering
    \includegraphics[width=\columnwidth]{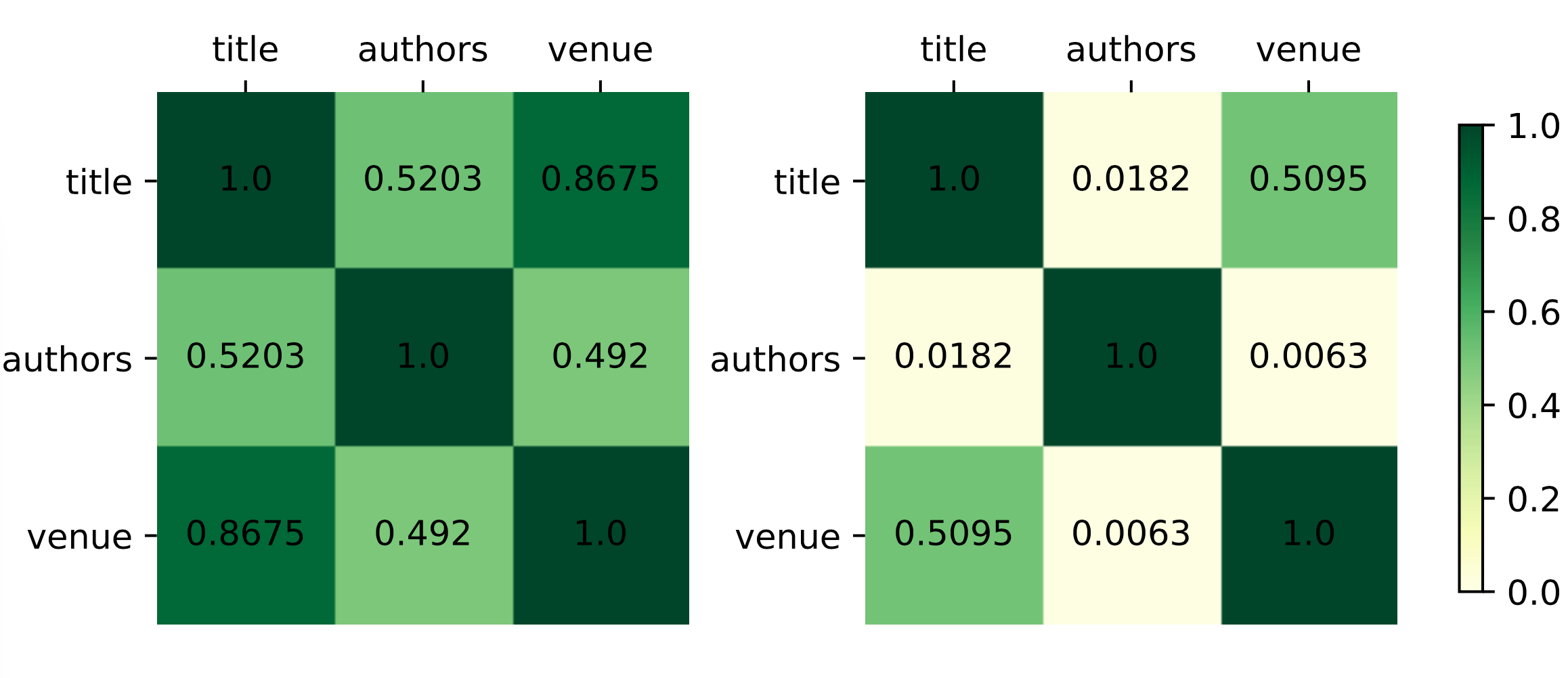}
    \caption{Pairwise cosine similarity of column embeddings: (a) \textit{WTE} embeddings; (b) \textit{\sysname-WTE} embeddings.}
    \label{fig:heatmap}
  \end{figure}

  To illustrate the suppression effect of contrastive learning, we compare heatmaps of pairwise cosine similarity of column embeddings encoded by \wte and \ourwte respectively. Consider the three text columns of the first table in Figure~\ref{fig:unionable_table_example}. As shown in Figure~\ref{fig:heatmap}(a), the pairwise cosine similarity of \wte embeddings is mostly above 0.5. There is a very high similarity (0.87) between the "title" column and the "venue" column and they will be mistakenly viewed as union-able. But this is not an issue for \ourwte embeddings as shown in Figure~\ref{fig:heatmap}(b) where the pairwise similarity between different columns are much lower (below 0.51) and the LSH index will not return the "venue" column as a union-able candidate of the "title" column.

  In the next set of experiments, we consider three syntactic measures used by \dddl and evaluate how much they can augment our embedding measures.
  \begin{enumerate}
    \item Name ($N$): Jaccard similarity between q-gram sets of attribute names.
    \item Value ($V$): Jaccard similarity between the TF-IDF sets of attribute values.
    \item Format ($F$): Jaccard similarity between regular expression sets of attribute values.
  \end{enumerate}

  \textbf{Experiment 2: Effectiveness of the ensemble of \sysname model variants and syntactic measures.} Figure~\ref{fig:exp2_effectiveness}(a) and (b) show the precision and recall of the ensemble of \sysname embedding models and syntactic measures on \sysname and TUS-Small datasets respectively. We consistently observe from both datasets that adding syntactic measures can further enhance the performance. In particular, name ($N$) and value ($V$) similarity are most effective syntactic measures. Around the average answer size of the \sysname dataset ($k=40$), $N$ and $V$ together raise up the precision and recall of \ourfasttext by nearly $20\%$, of \ourwte by $10\%$, and of \ourlm by over $5\%$. Similarly, around the average answer size of the TUS-Small dataset ($k=170$), there is an increase of about $10\%$ in both precision and recall for \ourfasttext, about $5\%$ for \ourwte, and more than $10\%$ for \ourlm.

\begin{figure}[!ht]
    \centering
    \minipage{0.49\linewidth}
      \includegraphics[width=\textwidth]{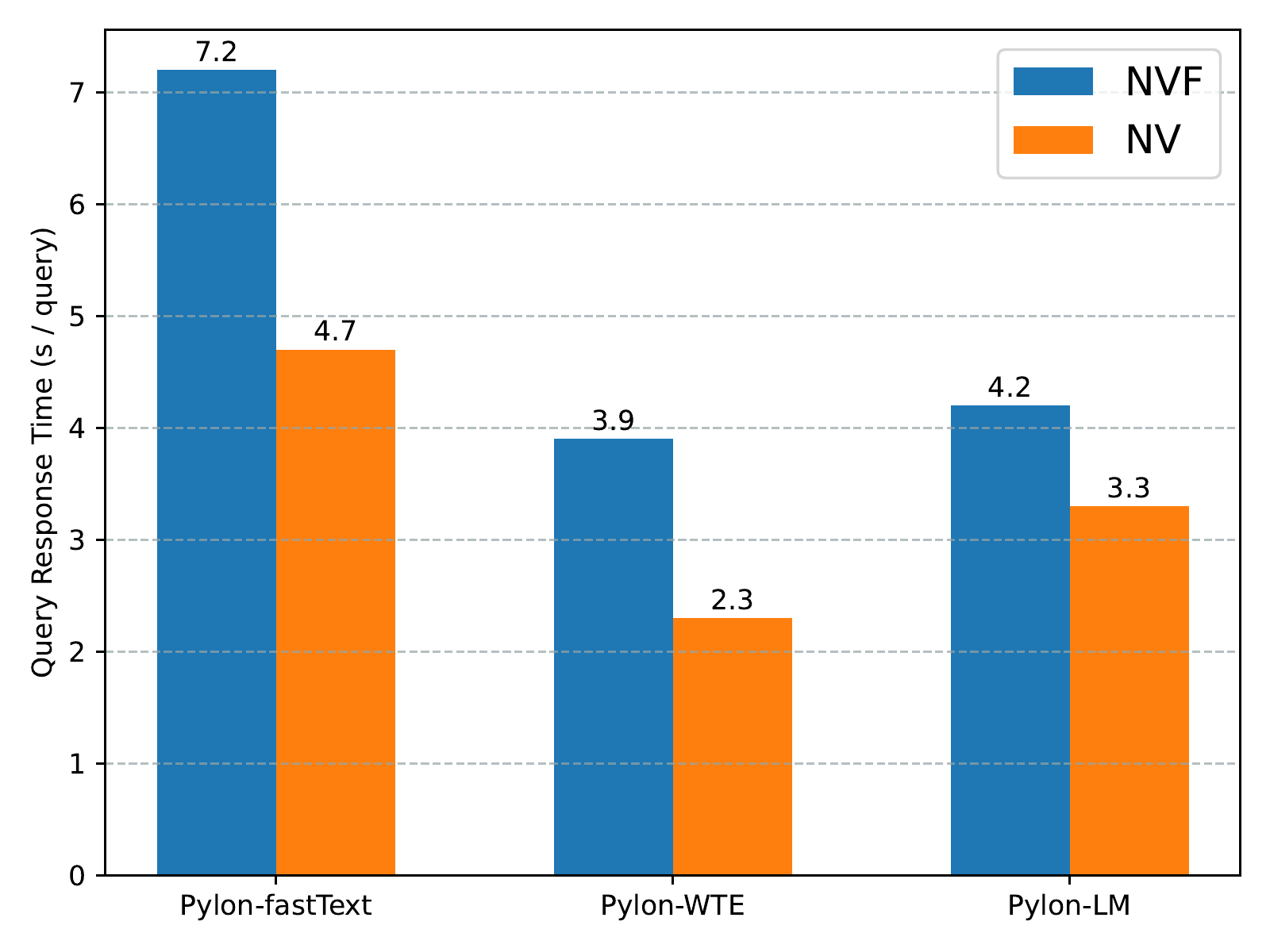}
      \caption*{(a) \sysname dataset}
    \endminipage\hfill
    \minipage{0.49\linewidth}
      \includegraphics[width=\textwidth]{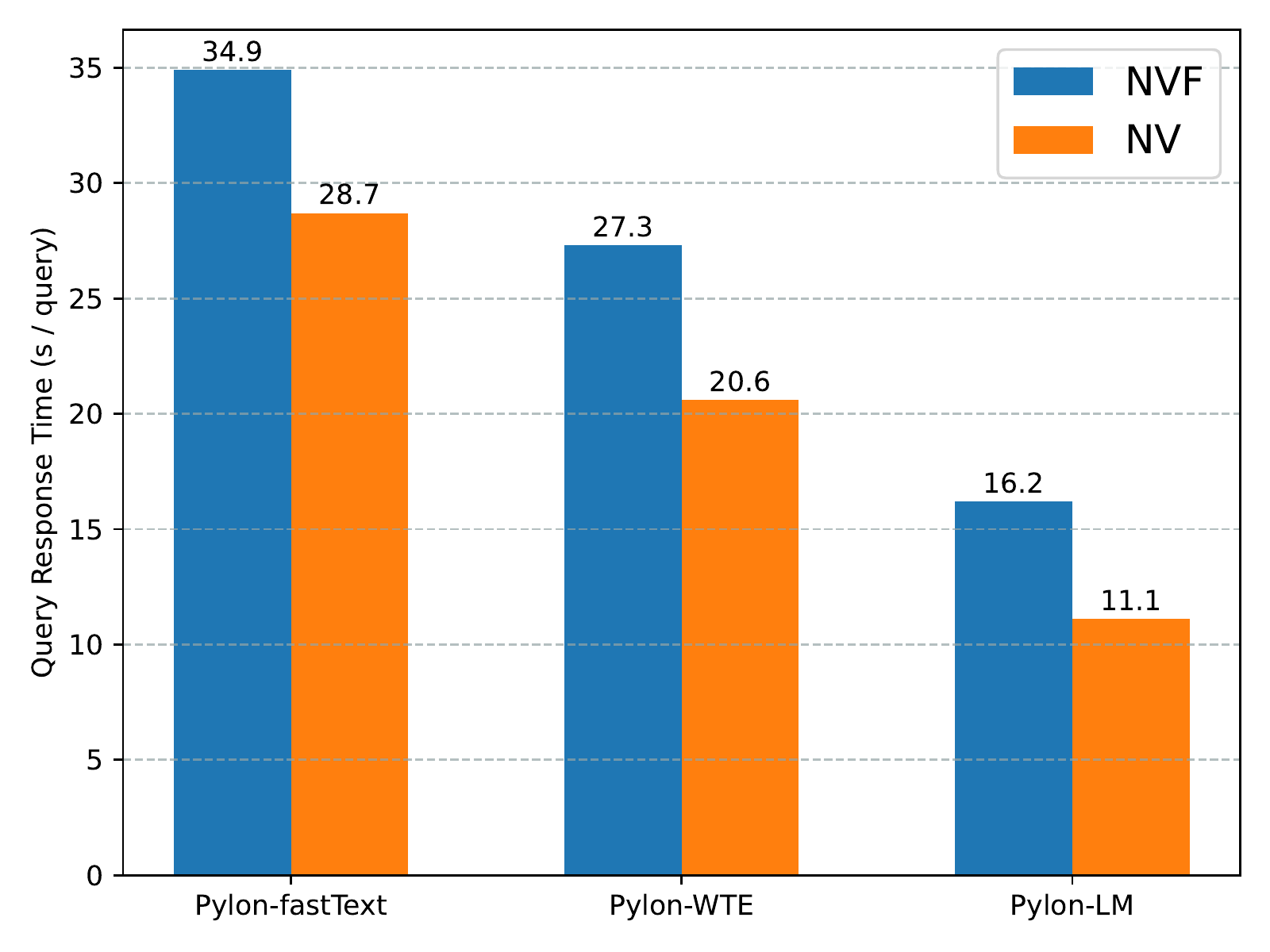}
      \caption*{(b) TUS-Small dataset}
    \endminipage
    \caption{Comparison of query response time between including and excluding the format measure.}
    \label{fig:exp2_efficiency}
  \end{figure}

  We also observe that adding additional format measure ($F$) hurts the performance (notably on the \sysname dataset and slightly on TUS-small). This is because tables in the \sysname dataset are mostly from disparate sources and so the value format tends to be inconsistent across tables whereas tables in TUS-Small are synthesized from only 8 base tables and it is much more likely for many tables to share format similarity. Even worse, including format index imposes non-trivial runtime cost (see figure~\ref{fig:exp2_efficiency}). For example, compared to model \textit{\sysname-WTE-NV}, the query response time of \textit{\sysname-WTE-NVF} (with the extra format measure) surges by $66.7\%$ on the \sysname dataset and by $32.2\%$ on TUS-Small.
    
  \begin{figure*}[!ht]
    \centering
    \minipage{0.33\textwidth}
      \includegraphics[width=\linewidth]{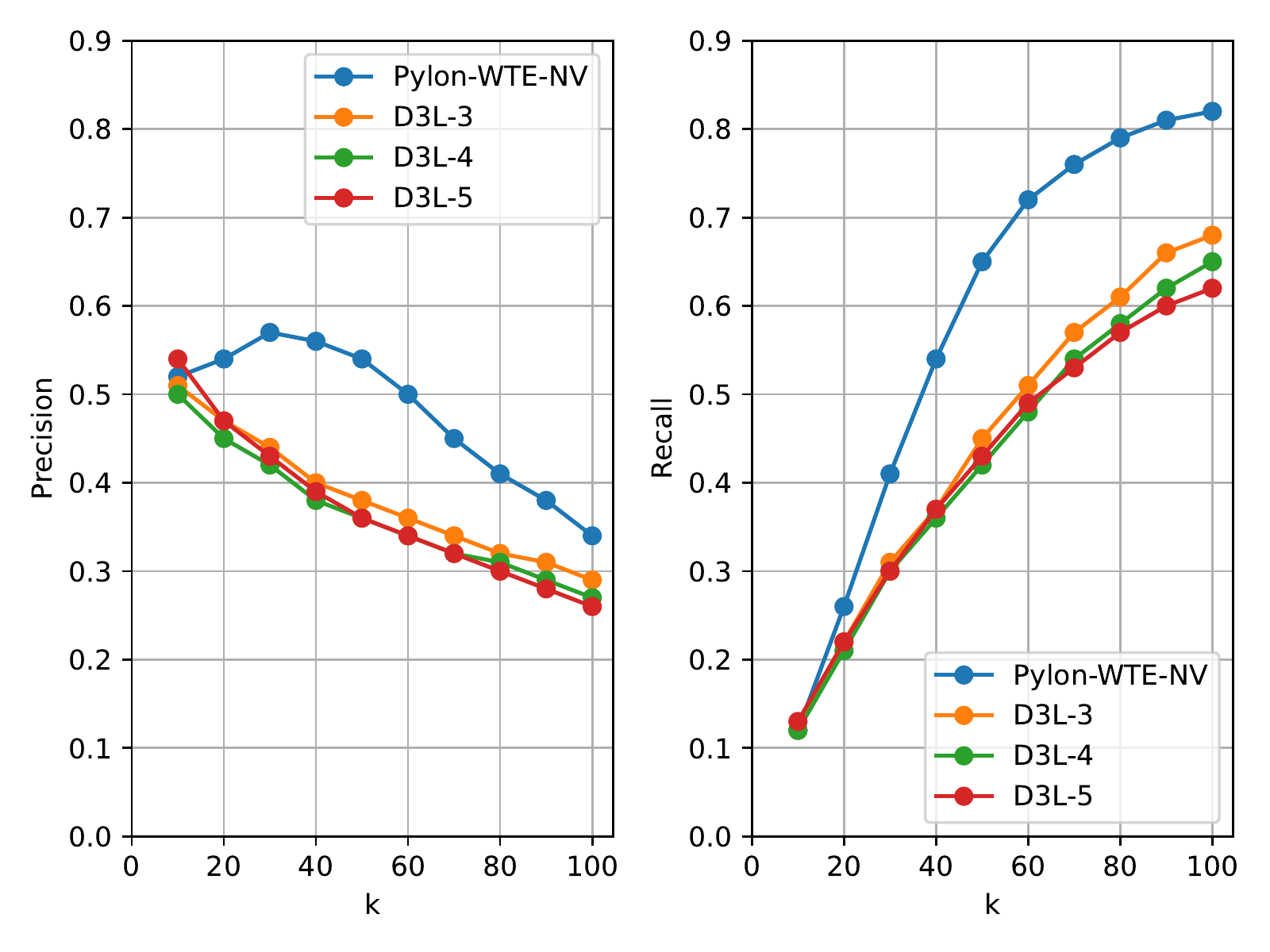}
      \caption*{(a) \sysname dataset}
    \endminipage\hfill
    \minipage{0.33\textwidth}
      \includegraphics[width=\linewidth]{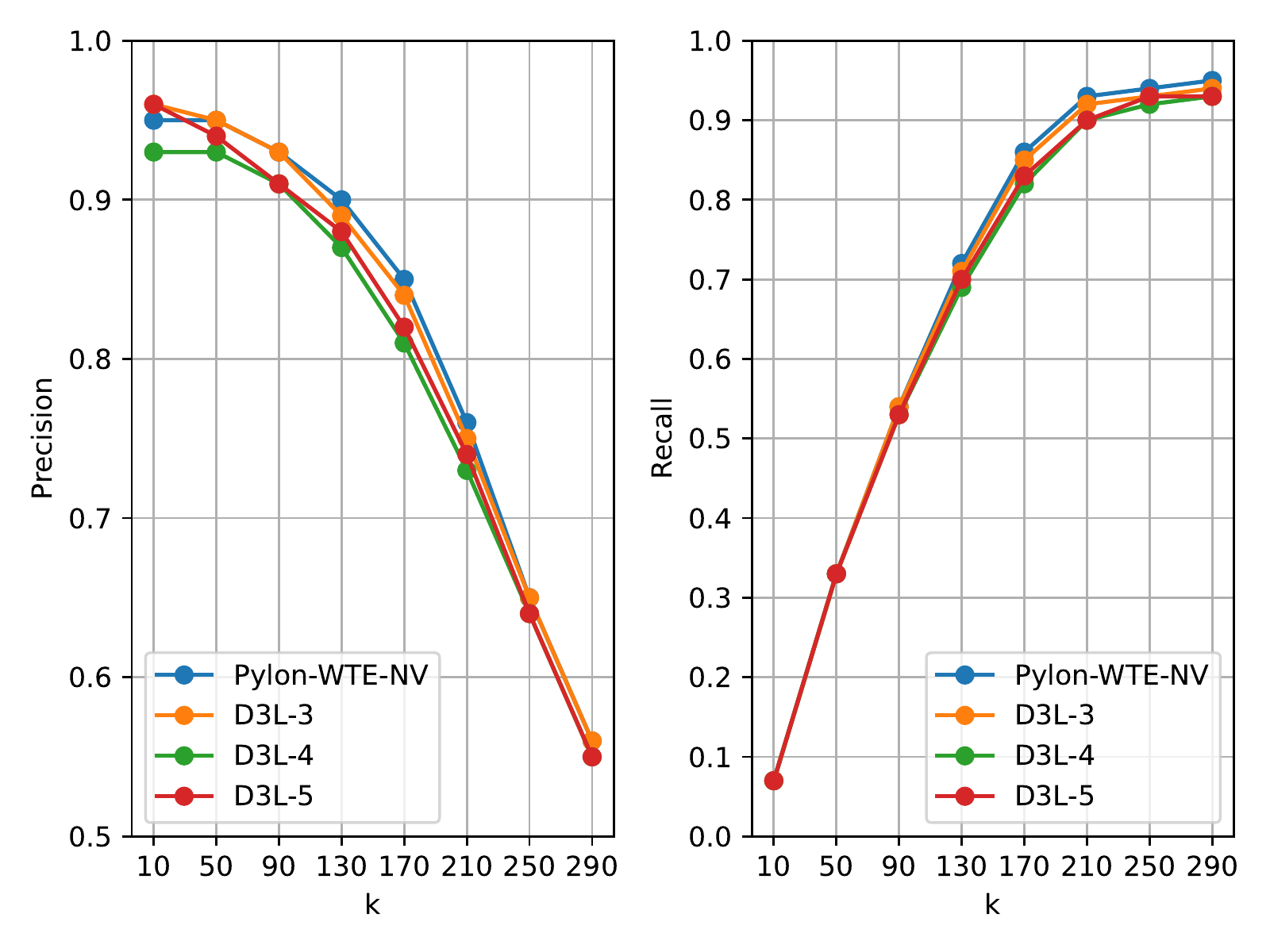}
      \caption*{(b) TUS-Small dataset}
    \endminipage\hfill
    \minipage{0.33\textwidth}
      \includegraphics[width=\linewidth]{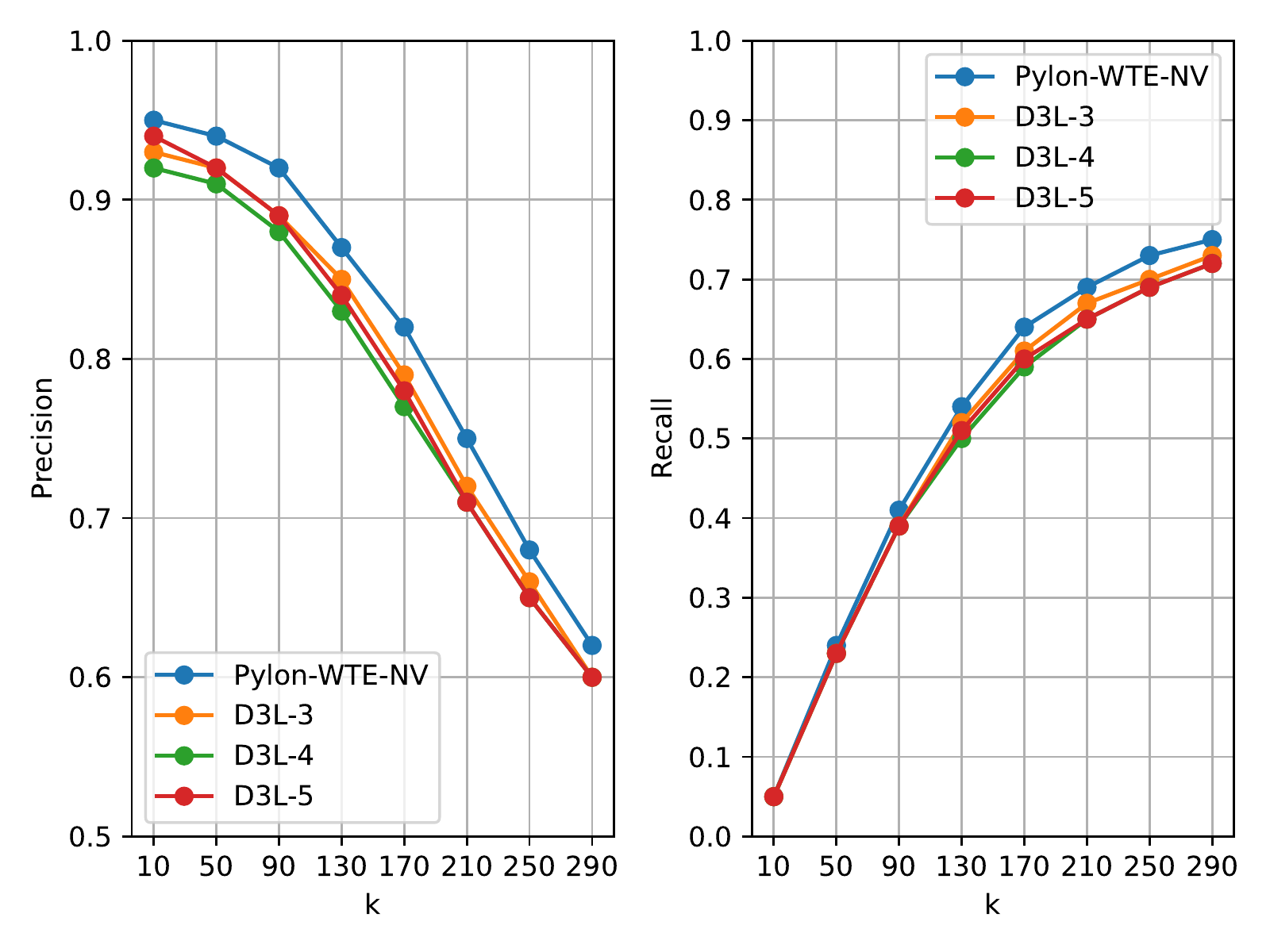}
      \caption*{(c) TUS-Large dataset}
    \endminipage
    \caption{Comparison of precision and recall between $\mathit{D^{3}L}$ instances and our best model \textit{\sysname-WTE-NV}.}
    \label{fig:exp3_effectiveness}
  \end{figure*}
      
  \begin{figure*}[!ht]
    \centering
    \minipage{0.33\textwidth}
      \includegraphics[width=\linewidth]{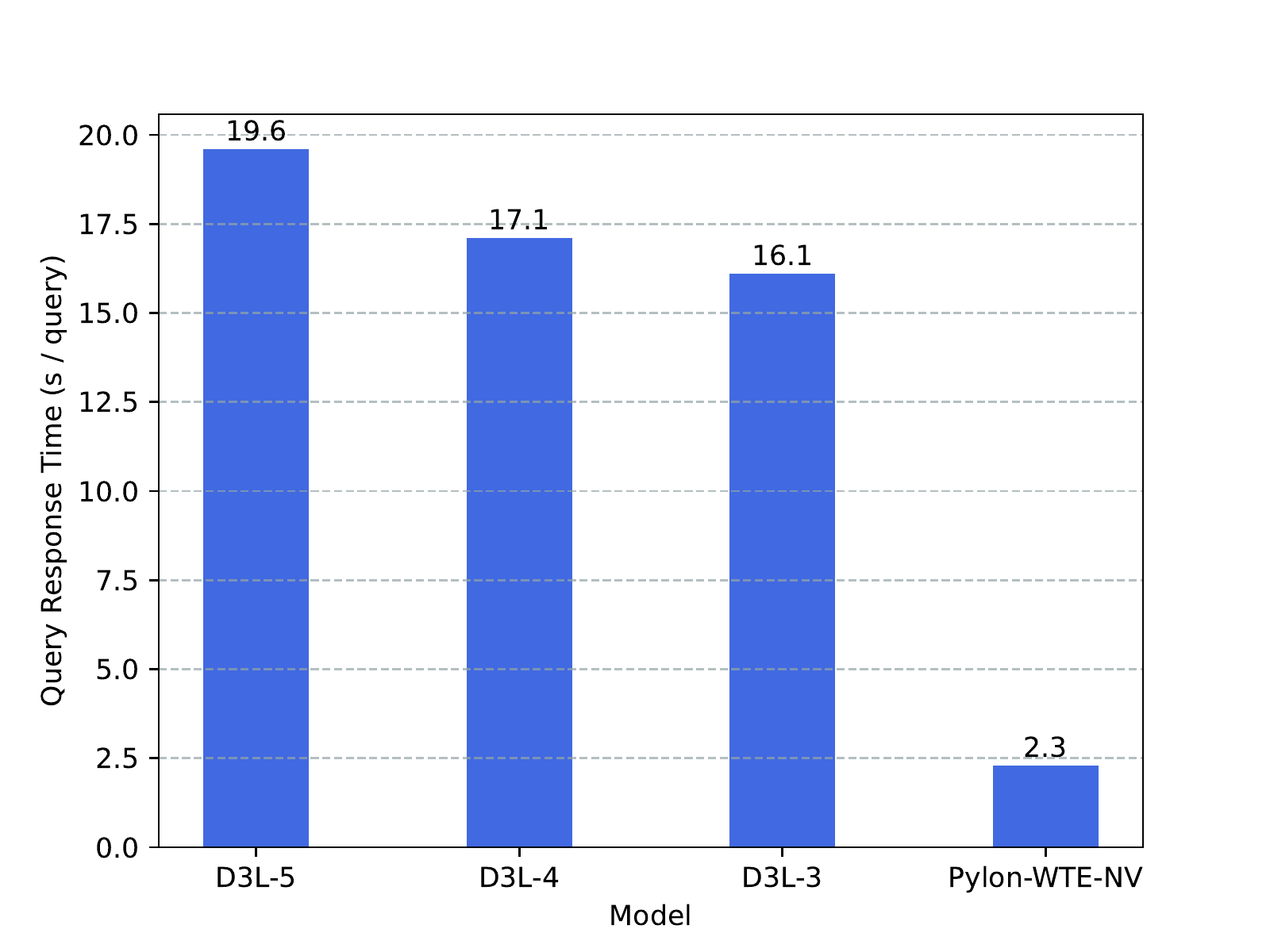}
      \caption*{(a) \sysname dataset}
    \endminipage\hfill
    \minipage{0.33\textwidth}
      \includegraphics[width=\linewidth]{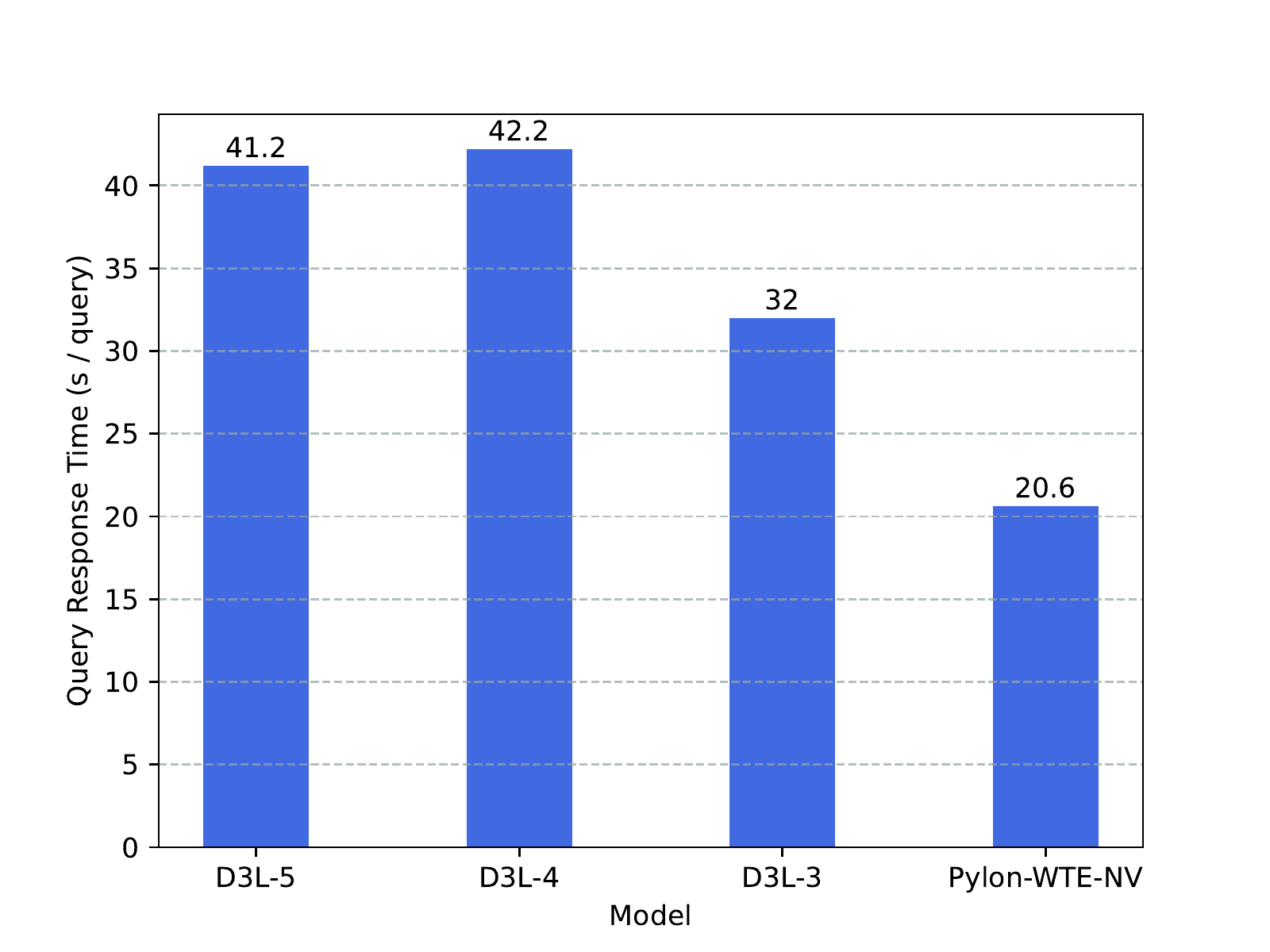}
      \caption*{(b) TUS-Small dataset}
    \endminipage\hfill
    \minipage{0.33\textwidth}
      \includegraphics[width=\linewidth]{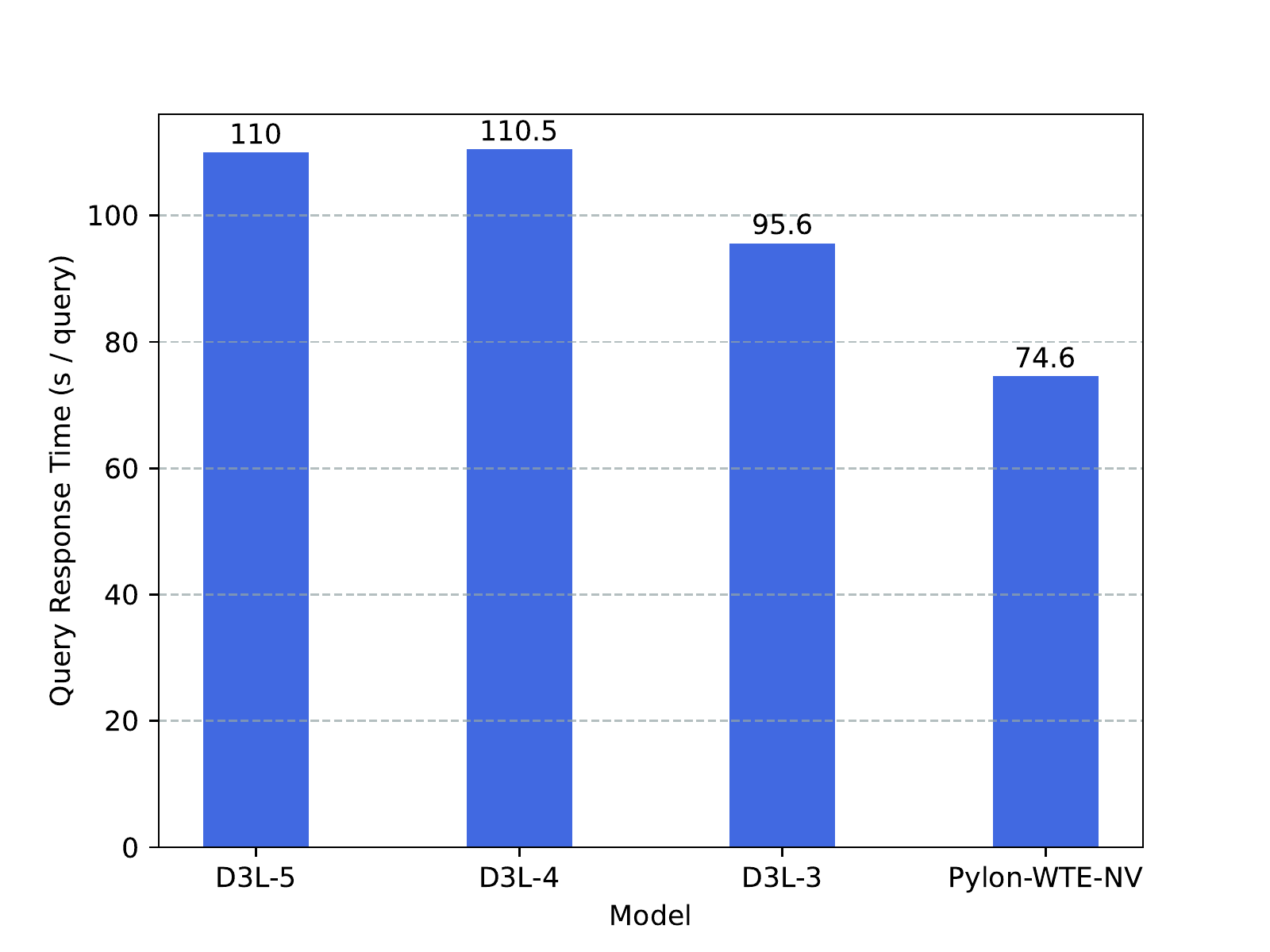}
      \caption*{(c) TUS-Large dataset}
    \endminipage
    \caption{Comparison of query response time between $\mathit{D^{3}L}$ instances and \textit{\sysname-WTE-NV}.}
    \label{fig:exp3_efficiency}
  \end{figure*}
    
  Finally, we compare our best-performing model \textit{\sysname-WTE-NV} with the state-of-the-art \dddl. As \textit{\sysname-WTE-NV} does not use format and domain measures in \dddl, for fair comparison, we consider three versions of \dddl. We refer to the full version of \dddl as \dddl-5, the one without the format measure as \dddl-4, and the one without format and domain measures as \dddl-3.
  
  \textbf{Experiment 3: Comparison of effectiveness and efficiency between our best model and \dddl.} Figure~\ref{fig:exp3_effectiveness} shows the performance of \textit{\sysname-WTE-NV} and three \dddl variants on \sysname, TUS-Small, TUS-Large datasets respectively. Around the average answer size ($k=40$) of the \sysname dataset,  \textit{\sysname-WTE-NV} is around $15\%$ better than the strongest \dddl instance (i.e., \dddl-3) in both precision and recall. \textit{\sysname-WTE-NV} performs much better than \dddl in this case because our embedding model using contrastive learning is optimized for indexing/search data structure and is trained on a dataset of a distribution similar to the test set and thus can capture more semantics than the off-the-shelf \fasttext embedding model used in $\mathit{D^{3}L}$.
    
  On TUS-Small and TUS-Large, we observe all instances have relatively competitive performance while \textit{\sysname-WTE-NV} performs marginally better compared to all \dddl variants. On TUS-Small, around the average answer size ($k=170$), \textit{\sysname-WTE-NV} is $2\%$ better than \dddl-3 and $5\%$ better than \dddl-5 in both precision and recall. On TUS-Large, around the average answer size ($k=290$), \textit{\sysname-WTE-NV} is more than $2\%$ better than $\mathit{D^{3}L}$ variants in both metrics. The smaller performance gap is due to the synthetic nature of TUS benchmark where most of union-able tables are generated from the same base table and share common attribute names and many attribute values. So syntactic measures ($N$ and $V$) can capture most of similarity signals and obtain high precision and recall even without support of semantic evidence.
    
  Additional to the performance gain, the biggest advantage of \textit{\sysname-WTE-NV} is the fast query response time. On the \sysname dataset, our model is nearly 9x faster than the full version \dddl-5 and 7x faster than \dddl-3. Even on TUS-Small and TUS-Large, which are datasets of a different data distribution (open data tables), we still save runtime by $44\%$ and $32\%$ respectively compared to \dddl-5, and by $35.5\%$ and $21.9\%$ respectively compared to \dddl-3.

\begin{figure*}[!ht]
    \centering
    \minipage{0.45\textwidth}
      \includegraphics[width=\linewidth]{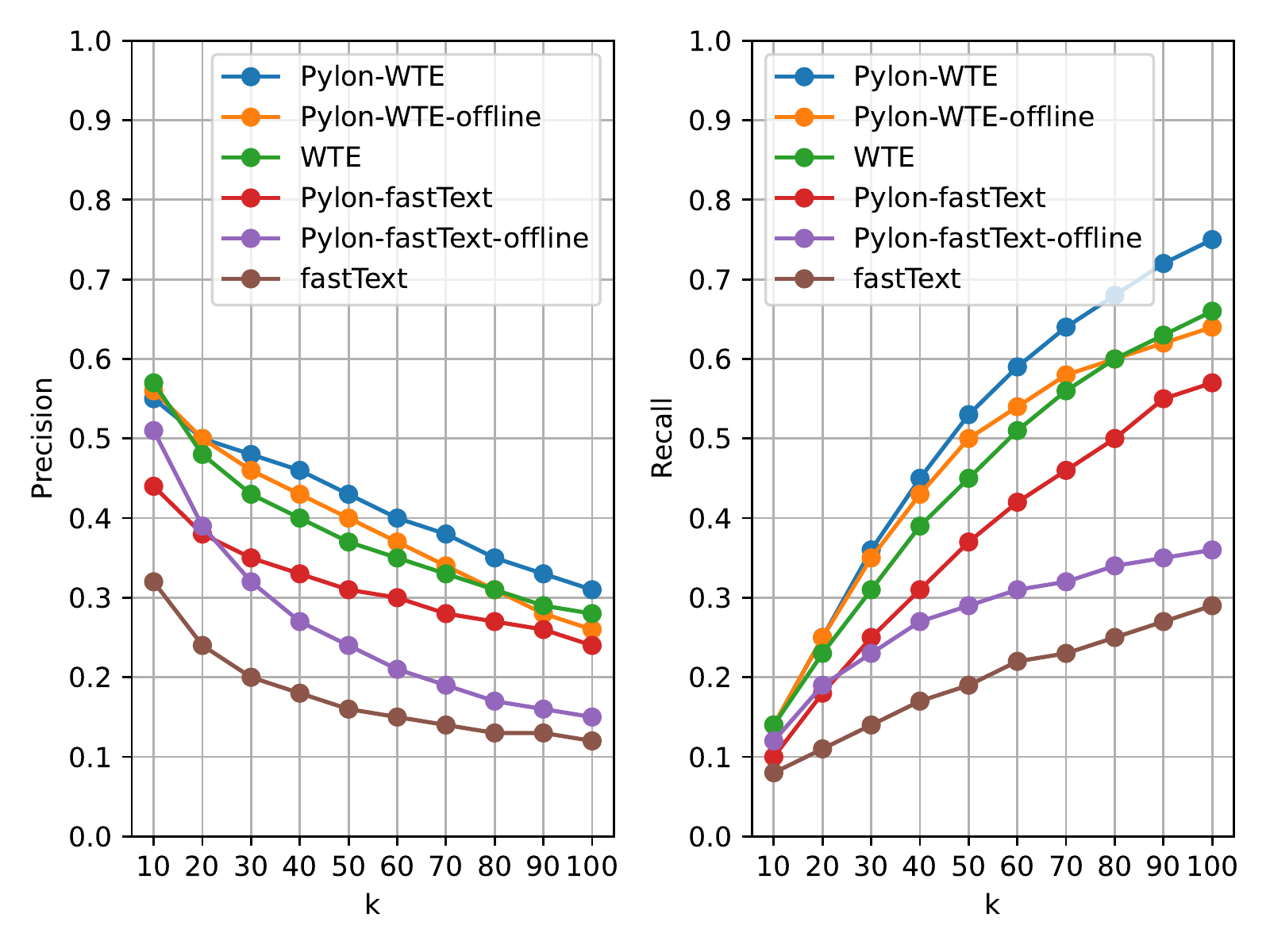}
      \caption{Top-k precision and recall of 6 embedding measures on the \sysname dataset.}
      \label{fig:exp4_pylon_performance}
    \endminipage\hfill
    \minipage{0.45\textwidth}
      \includegraphics[width=\linewidth]{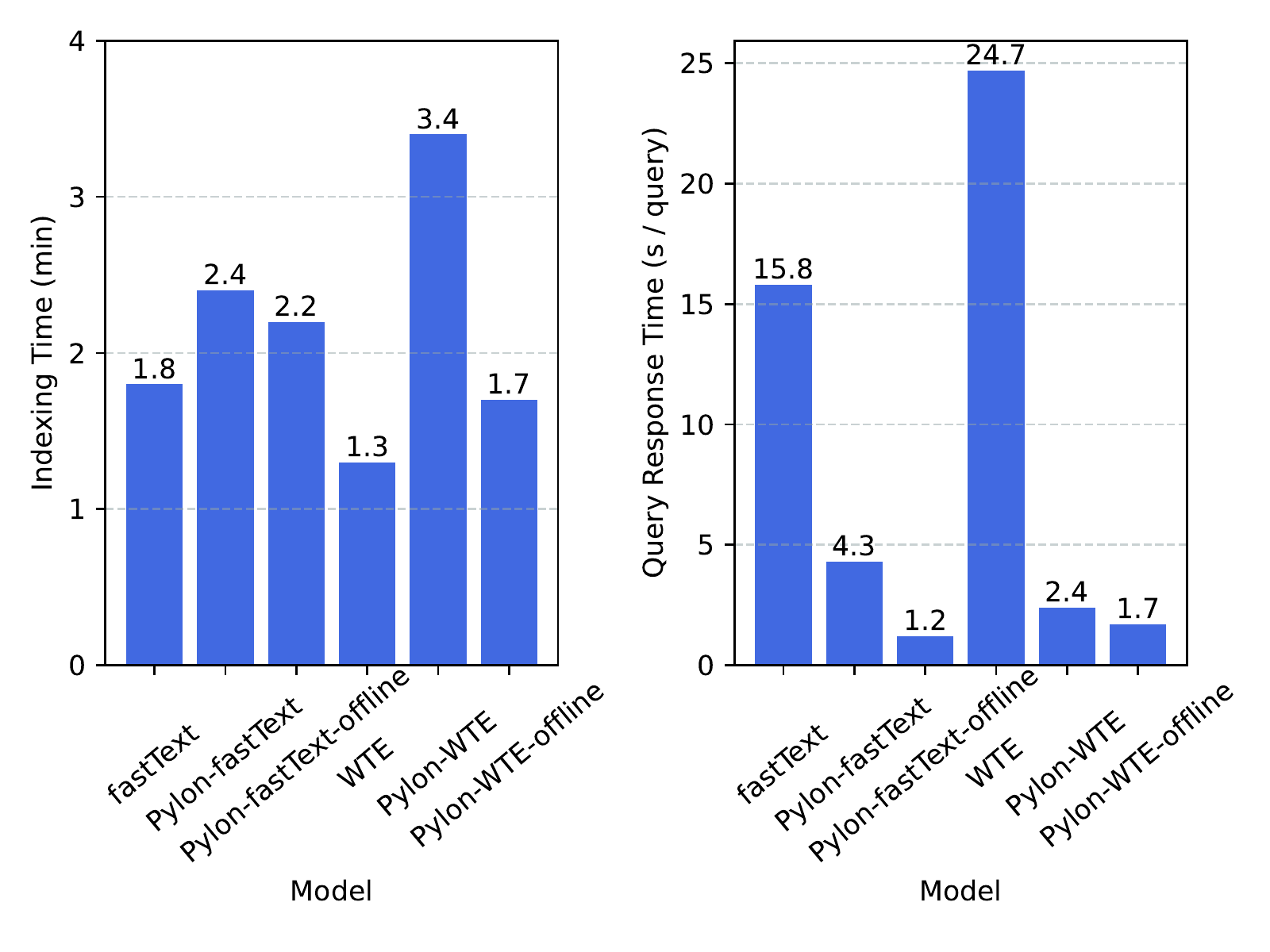}
      \caption{Indexing time and query response time on the \sysname dataset.}
      \label{fig:exp4_pylon_efficiency}
    \endminipage
  \end{figure*}

  \textbf{Experiment 4: Effectiveness and efficiency of \sysname model variants from the offline training data construction strategy.} Figure~\ref{fig:exp4_pylon_performance} shows the precision and recall of 4 \sysname variants from two training data construction strategies and their baselines. On the \sysname dataset, around the average answer size ($k=40$), two \sysname models from the alternative data construction strategy, \textit{\sysname-WTE-offline} and \textit{\sysname-fastText-offline}, retain strong performance and outperform the corresponding baseline by $3\%$ and $9\%$ respectively. Note that \sysname models derived from the sampling data construction strategy have consistently better performance as $k$ increases. We also observe a similar trend on the TUS benchmark while the performance gap of all instances is smaller. 
    
  As shown in figure~\ref{fig:exp4_pylon_efficiency}, both new models are efficient in indexing time and query response time. Compared to the corresponding baseline, \textit{\sysname-WTE-offline} is 12x faster and \textit{\sysname-fastText-offline} is 14.5x faster in query response time. Again, this significant speedup demonstrates the value of considering characteristics of indexing/search data structure in the training of embedding models and the distinguishing power of contrastive learning, which enables the LSH index to work more effectively with embeddings.   
    
\subsection{Discussion}
  We also leave and discuss a few clues for future extensions. 

  \textbf{Alternative Contrastive Loss.} While the loss function used in this project is an effective one, it is not the only feasible training objective for self-supervised contrastive learning. For example, triplet loss~\cite{schroff2015facenet} considers a triplet $(x, x^{+}, x^{-})$ as a training example where $x$ is an input, $x^{+}$ is a positive sample (belonging to the same class as $x$ or semantically similar to $x$) and $x^{-}$ is a negative sample. Additionally, what considers as negative examples and "hardness" of negative examples are also interesting aspects to explore.

  \textbf{Verification of Column Union-ability.} Besides quantitative evaluation, we also manually inspect results of a few queries for each dataset. We observe that in some correct table matches, there are misalignment of union-able columns. To mitigate this issue, we consider that progress in semantic column type prediction~\cite{zhang2020sato, suhara2022annotating} can be beneficial for verifying the union-ability of columns as a post-processing step.

%% file: sections/05_relatedWork.tex
\section{Related Work}
  Our work is most related to data integration on the web and data discovery in enterprise and open data lakes. 

  \textbf{Web Table Search.} \cite{cafarella2009data} presents OCTOPUS that integrate relevant data tables from relational sources on the web. OCTOPUS includes operators that perform a search-style keyword query over extracted relations and their context, and cluster results into groups of union-able tables using multiple measures like TF-IDF cosine similarity. ~\cite{yakout2012infogather} defines three information gathering tasks on Web tables. The task of augmentation by example essentially involves finding union-able tables that can be used to fill in the missing values in a given table. Their Infogather system leverages indirectly matching tables in addition to directly matching ones to augment a user input. ~\cite{Sarma12} formalizes the problem of detecting related Web tables. At the logical level, the work considers two tables related to each other if they can be viewed as results to queries over the same (possibly hypothetical) original table. In particular, one type of relatedness they define is \textit{Entity Complement} where two tables with coherent and complementary subject entities can be unioned over the common attributes. This definition requires each table to have a subject column of entities indicating what the table is about and that the subject column can be detected. Following the definition, the work captures entity consistency and expansion by measuring the relatedness of detected sets of entities with signals mined from external ontology sources. Finally, they perform schema mapping of two complement tables by computing a schema consistency score made up of the similarity in attribute names, data types, and values. 
  
  \textbf{Data Discovery in the Enterprise}. ~\cite{DBLP:conf/sigmod/FernandezAMS16} identifies data discovery challenges in the enterprise environment. The position paper describes a data discovery system including enrichment primitives that allow a user to perform entity and schema complement operations. Building on top of the vision in ~\cite{DBLP:conf/sigmod/FernandezAMS16}, ~\cite{fernandez2018aurum} presents AURUM, a system that models syntactic relationships between datasets in a graph data structure. With a two-step process of profiling and indexing data, AURUM constructs a graph with nodes representing column signatures and weighted edges indicating the similarity between two nodes (e.g., content and schema similarity). By framing queries as graph traverse problems, AURUM can support varied discovery needs of a user such as keyword search and similar content search (which can be used for finding union-able columns and tables). ~\cite{fernandez2018seeping} further employs word embeddings in AURUM to identify semantically related objects in the graph.
  
  \textbf{Data Discovery in Open Data Lakes}. ~\cite{Nargesian18} defines the table union search problem on open data and decomposes it as finding union-able attributes. They propose three statistical tests to determine the attribute union-ability: (1) set union-ability measure based on value overlap; (2) semantic union-ability measure based on ontology class overlap; and (3) natural language union-ability measure based on word embeddings. A synthesized benchmark consisting of tables from Canadian and UK open data shows that natural language union-ability works best for larger $k$ in top-$k$ search. In the meantime, set union-ability is decent when $k=1$ for each query but vulnerable to value overlap in attributes of non-unionable tables, and semantic union-ability stays competitive to find some union-able tables for most queries despite incomplete coverage of external ontologies. The ensemble of three measures further improves the evaluation metrics. ~\cite{bogatu2020dataset} adopts more types of similarity measures based on schema- and instance-level fine-grained features. Without relying on any external sources, their \dddl framework is shown effective and efficient on open data lakes. E{\scriptsize MB}DI~\cite{cappuzzo2020creating} proposes a graph model to capture relationships across relational tables and derives training sequences from random walks over the graph. They further take advantage of embedding training algorithms like \textit{fastText} to construct embedding models. Their relational embeddings demonstrate promising results for data integration tasks such as schema matching and entity resolution. SANTOS~\cite{khatiwada2022santos}, a very recent work, leverages the relationship of pairs of attributes for table union search. However, it relies on the existence of a relationship and the coverage of the relationships in a knowledge base. Although SANTOS also draws on the data lake itself to discover new relationships using the co-occurrence frequency of attribute pairs, it particularly misses rare yet important relationships. 
  
  For a broader overview of the literature, we refer readers to the survey of dataset search~\cite{chapman2020dataset}.



%% file: sections/06_conclusion.tex
\section{Conclusion}
  In this work, we formulate the table union search problem as an unsupervised representation learning and embedding retrieval task. We present \sysname, a self-supervised contrastive learning framework that models both data semantics and characteristics of indexing/search data structure. We also demonstrate that our approach is both more effective and efficient compared to the state-of-the-art. 
  